# Alternative Blockmodelling

Oscar Fabian Correa Guerrero

Student number: 594842

Supervisors: Dr. Jeffrey Chan and Dr. Vinh Nguyen

Number of credit points: 75

Current research unit: COMP60002 Computer Science Research Project

# Abstract

Many approaches have been proposed to discover clusters within networks. Community finding field encompasses approaches which try to discover clusters where nodes are tightly related within them but loosely related with nodes of other clusters. However, a community network configuration is not the only possible latent structure in a graph. Core-periphery and hierarchical network configurations are valid structures to discover in a relational dataset.

On the other hand, a network is not completely explained by only knowing the membership of each node. A high level view of the inter-cluster relationships is needed. Blockmodelling techniques deal with these two issues. Firstly, blockmodelling allows finding any network configuration besides to the well-known community structure. Secondly, blockmodelling is a summary representation of a network which regards not only membership of nodes but also relations between clusters.

Finally, a unique summary representation of a network is unlikely. Networks might hide more than one blockmodel. Therefore, our proposed problem aims to discover a secondary blockmodel representation of a network that is of good quality and dissimilar with respect to a given blockmodel.

Our methodology is presented through two approaches, *(a) inclusion of cannot-link constraints* and *(b) dissimilarity between image matrices*. Both approaches are based on non-negative matrix factorisation *NMF* which fits the blockmodelling representation. The evaluation of these two approaches regards quality and dissimilarity of the discovered alternative blockmodel as these are the requirements of the problem.

# Contents

# List of Figures

# List of Tables

# Chapter 1

# Introduction

When an individual is confronted to raw data, among her first reactions would be trying to make sense of them somehow. Getting patterns from data could end being a painful task especially if the amount of data is massive and/or these data has many features. Furthermore, appreciation of the data and their grouping might be correct for certain users, but it may not make sense for others. Therefore, thinking about the existence of a unique grouping of the data is not adequate and even limiting.

**Extensive research has been carried out on alternative clusterings in non-relational data** [1-7]. When datasets have many features, it is tempting to think that data objects can be clustered regarding a different subset of features each time an alternative solution is searched. For instance, a document repository may contain millions of documents where each of them has metadata and topic features. Then, an algorithm can find a candidate clustering of these documents by taking into account the metadata in first place, and it can also find an alternative clustering by regarding the topic-based features in second place.

On the other hand, **with respect to relational data, community finding has been a widely studied field** which focuses on discovering hidden structures within a network. These structures are called communities and consist of nodes that are tightly related between them but loosely related with the ones of other communities. Nonetheless, unlike non-relational data, most algorithms in community finding do not offer an alternative solution even when another clustering of communities could be more suitable to a determined user.

Moreover, **a totally different hidden structure can be found since community structure is not the only possible latent configuration**. In fact, many real networks follow a power-law degree distribution –at least asymptotically- where there is a small number of central or "core" nodes with lots of connections and a heavy tail of "periphery" nodes with a few connections. This is the case of social networks, e.g. Twitter®, which has a limited number of very active users with lots of friends and a considerable amount of passive –not as active as the former- users with fewer connections. Thus, a community structure would be difficult to discover in this kind of networks.

Discovering an alternative community clustering or a completely different network structure may be more convenient for certain purposes. A user would rather have more options when she is facing a clustering problem, especially when it is related with a decision-making process. It should be said that **the expected alternative solution must be as good as the initial proposed clustering** that is, it ought to show a coherent division of the network. However, **a trade-off situation might arise between quality and dissimilarity since as dissimilarity increases quality could be affected and vice versa**.

Besides to the possibility of discovering different network structures, an alternative clustering finding approach in graphs might offer more information. Knowing the structure and cluster membership of the nodes can yield a good understanding of the network, yet it is incomplete. Relationships between clusters help learn how clusters interact, what role each of them plays within the whole. This is deeply true in social network analysis where relations –block within social analysis jargon- explicitly exist apart of clusters –position[1] is the social term used instead of clusters. For example, within a trade network, consumers and producers are sound candidate positions but there also exists a marked consumer-producer relationship which is needed to explain the network completely.

Therefore, this thesis presents a novel approach which overcomes these issues by combining blockmodelling techniques [8] and an efficient and well-proved optimisation approach called non-negative matrix tri-factorisation [9-11]. Firstly, through blockmodelling, it is suitable to represent an entire network by two basic matrices: position membership matrix and an image matrix, i.e. the matrix which shows the relations between clusters. Moreover, blockmodelling manages to summarise a complex network through a smaller comprehensible representation that can encompass different network structures such as communities, core-periphery, and so on. Thus, community finding is a special case of blockmodelling where the image matrix is a diagonal matrix. Secondly, the clustering process can be regarded as a non-convex optimisation problem where a local optimum might be found by decomposing the network adjacency matrix[2] into three sub-matrices factors: position membership matrix –twice- and image matrix. This process is carried out iteratively through multiplicative update rules [11, 12]. Finally, the proposed objective functions enclose both quality and dissimilarity of the alternative blockmodel.

The rest of this chapter presents brief background information of relevant terms within the context of this work. Finally, it also highlights the contributions of this research, yet future work is possible.

## 1.1. Background

In this section, important related concepts are briefly explained. *Graphs, clustering, blockmodelling* and *optimisation* are terms used intensively throughout this thesis and thus, they need to be explained. Furthermore, contributions of this work are presented.

*Graphs* consist of nodes, edges and their attributes. A node is also known as vertex and may represent any entity within a network, e.g. a Facebook® user, a protein within a protein interaction network or a computer in a LAN[3]. A node is connected to other nodes through edges. Both, nodes and edges might have attributes. A node attribute could be its name in the case of a Facebook® user, whereas an edge attribute could be the level of friendship between two Facebook® users.

---

[1] **Terms *position* and *cluster* can be used interchangeably.**
[2] Adjacency matrix $A \in \mathbb{R}^{n \times n}$ is the representation of a network of $n$ nodes where each cell $a_{ij} \in \mathbb{R}$ corresponds to the weight of the edge between the $i^{th}$ and $j^{th}$ nodes.
[3] LAN = Local Area Network

*Clustering* is an application of data mining whose aim is to group similar entities within a dataset. In order to complete its task, clustering does not count on user supervision. That is, there is no extra information which would guide its task. However, a recent research field called semi-supervised clustering can lean on the existence of a small number of labelled entities which may help guide the clustering process per se. Furthermore, similarity between entities is an application-dependent measure –the reader will see that a special similarity criteria is regarded within this work.

*Blockmodelling* (See Figure 1-1) aims to summarise the whole structure of the network – adjacency matrix- through two matrices: (a) *position membership matrix*, and (b) *image matrix*. In the position membership matrix, each row corresponds to a node and each column to a cluster. The resulting cells have either one or zero[4], depending on whether the node belongs or not to the cluster. On the other hand, the image matrix has the clusters as both rows and columns, and the cells represent the relation between clusters. Each cell in the image matrix is known as a *block* and it has a value of one if there is a relation between clusters or zero otherwise[5]. Both matrices form a blockmodel [8].

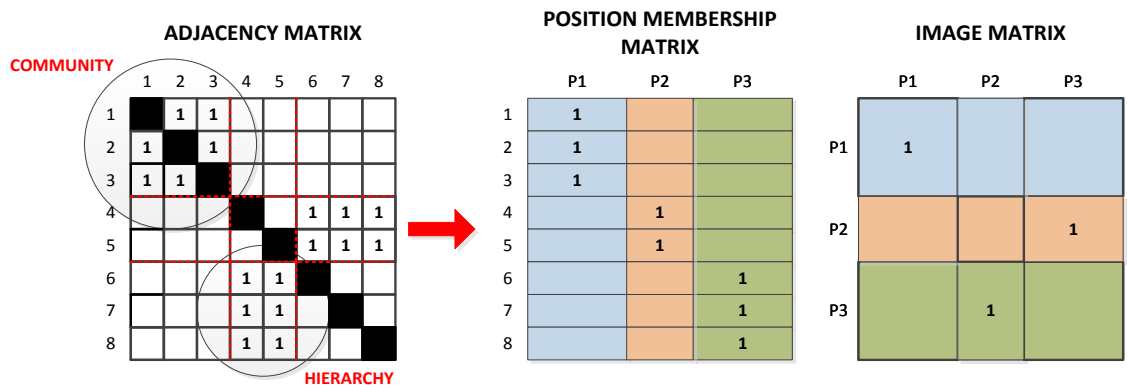

**Figure 1-1: Blockmodelling**

An adjacency matrix (left) can be summarised through two matrices (right): *position membership matrix* and *image matrix*. In the position membership matrix, each position corresponds to a column and each node to a row. In the image matrix, each cell is called *block*. A *block* shows if a relation between two positions does exist (1) or not (0). This is a distinctive characteristic of blockmodelling since it allows a better comprehension of the whole network.

*Optimisation* is a field in mathematics and computer science where the best element is searched within many possible alternatives and it may be bounded by one or more constraints.

---

[4] The values of the position membership matrix are not actually 1 or 0, but within [0-1] range. This is because the matrix factorization process is an approximation of the original adjacency matrix. More details in Chapter 3.
[5] Image matrix will also have values within [0-1] range. More details in Chapter 3.

Chapter 3 shows our approaches that count on optimisation to find a local minimum for the proposed objective function.

## 1.2. Contributions of the Thesis

This section presents the contributions of this thesis:

### 1.2.1. Proposal of the Alternative Blockmodelling problem in relational data

As far as we know, alternative clustering of a graph through blockmodelling techniques has not been introduced as a problem in the field. There has been extensive research in alternative clustering upon non-relational data [1-7]. There are also many algorithms to discover latent communities within networks [13-23]. However, alternative clustering of relational data in conjunction with its summary through blockmodel representation is a brand-new problem.

Thus, the proposed problem consists of not only finding an alternative clustering in a graph given an initial clustering but also revealing interrelations between its clusters through a blockmodel representation. Throughout this thesis, valuable characteristics related with blockmodels are presented.

Finally, the problem also requires discovering a different alternative blockmodel representation in relation with a given one, yet preserving a plausible division of the network.

### 1.2.2. Approach to discover a different and plausible alternative blockmodel representation with respect to a given one

We approach the problem based on the clustering potentiality that matrix factorisation has [24]. Moreover, matrix factorisation ends up to be closely linked to a blockmodel representation of a network since the original adjacency matrix is decomposed into two lower-rank matrices. These two matrices correspond to the *position cluster membership* and *image matrices* which are the basis of blockmodelling (See Figure 1-1.) Then, a candidate clustering will be the best approximation to the matrix factorisation. Therefore, the problem becomes an optimisation problem.

However, this part of the approach would only allow discovering a block model which could even be the same initial clustering. Then, the optimisation problem demands to be bounded somehow such that the discovered clustering is dissimilar to the given one. Hence, the objective function[6] requires taking into account the given clustering.

---

[6] Objective function is the equation to be optimized –minimised or maximised given certain constraints.

### 1.2.3. Insights into challenges and problems presented when approaching the proposed problem by using instance-based constraints

Our first approach focused on matrix factorisation and semi-supervised clustering by including instance-level constraints [25]. These constraints were built upon the given clustering in such a way that the discovered alternative blockmodel is dissimilar to it. Although, this approach looks promising since semi-supervised clustering has been successful by benefiting of the presence of very limited supervisory information –in this case, the given clustering-, it also presents some challenges. These challenges are especially related with their construction and identification.

Therefore, we offer some insights into the problems, challenges and future work within the application of instance-level constraints when finding a dissimilar clustering. Chapter 5 shows some conclusions about this approach, but more importantly its challenges and considerations.

# Chapter 2

# Literature Survey

Discovering an alternative blockmodel within a graph is related with alternative clustering in non-relational data, community finding within networks and blockmodelling techniques. Nonetheless, none of them strictly finds an alternative blockmodel representation of a network based on a given one. This chapter surveys these techniques and explains why they are different from the proposed problem.

## 2.1. Alternative clustering in non-relational data

Alternative clustering approaches in non-relational data are techniques which discover alternative groupings within multidimensional datasets, e.g. document repositories, protein databases, etc. Alternative clustering has been approached from different perspectives from which we survey two of them: information-theoretic-based and instance-level-constraint-based methods.

### 2.1.1. Information-theoretic approaches

Conditional Information Bottleneck –*CIB*- [6] is an information-theoretic approach that minimises the mutual information between the data objects and the cluster labels and maximises the mutual information between the features and cluster labels given the side information –initial clustering. According to [1], *CIB* is the most relevant work in alternative clustering field. However, *CIB* makes some assumptions when estimating the distribution of the random variables, i.e. data objects, features, cluster labels and side information.

In contrast to *CIB*, another information-theoretic approach called *minCEntropy* offers the possibility of finding a clustering or an alternative clustering without being provided a probability distribution [7]. *minCEntropy* works in an unsupervised manner which allows discovering a clustering, whereas *minCEntropy$^+$* and *minCEntropy$^{++}$*, semi-supervised versions of *minCEntropy*, find alternative clusterings with respect to a given clustering.

Similarly to *CIB, minCEntropy* is an objective-function-oriented approach which minimises the *conditional entropy* based on the intuition that entropy is the inverse of mutual information. By using mutual information, an algorithm tries to maximise the information between the data and the clustering. Then, clustering by using entropy will minimise it, which is the principle of this approach.

A very similar approach to *minCEntropy* is proposed in [4]. The authors called their method Non-linear Alternative Clustering with Information theory –*NACI*-. It is also an information-theoretic method that maximises the information between the clustering and the data objects while minimises the information between the sought alternative and the reference clustering.

*NACI* is oriented to find alternative clusterings on non-linear datasets, that is, it is not limited to spherically shaped clusterings. *minCEntropy* and *NACI* benefit from a kernel density estimation tool known as *Parzen windows* which allows them to estimate the probability densities of their random variables. In this way, the pitfall encountered in *CIB,* associated with its probability distribution assumption, can be avoided. However, both approaches need to be provided the standard deviation –width- $\sigma$ for the kernel function since they use a Gaussian kernel.

Clustering for Alternatives with Mutual Information –*CAMI*- is a neat and mathematically well founded information-theoretic approach which has an unsupervised nature [3]. Its objective function is conceived in such a way both clusterings are found at the same time, i.e. the reference and the alternative clusterings, through a Gaussian mixture where each cluster is a multivariate Gaussian distribution. Although *CAMI* looks similar to *CIB, CAMI* proved to be more accurate. The explanation resides on *CAMI*'s explicit quality assurance –likelihood maximisation- which *CIB* clearly lacks.

### 2.1.2. Instance-level constraint-based approaches

Besides to information-theoretic approaches, there are methods which encode the side information, i.e. the given clustering, as the so called instance-level constraints.

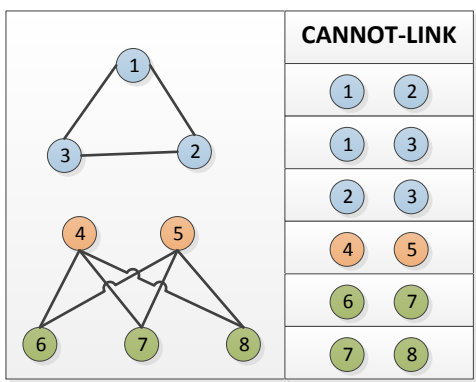

**Figure 2-1: Cannot link constraints**

The figure shows two subgraphs within the given clustering. In the first one (top,) one cluster is depicted with blue colour. From these three nodes, 3 cannot-link constraints were created and they are shown in the top of the table. Similarly, in the second subgraph, two clusters are depicted and their cannot-link constraints are also included in the table.

An instance-level constraint encapsulates the relationship between two data objects – instances- within the given clustering. Let $x_i$, $x_j\ where\ i \neq j$ be two instances in a dataset. If both instances belong to the cluster $c_k$, a **cannot-link** constraint is created. *Cannot-link* constraints identify instances that cannot-be-linked –cannot belong to the same cluster- in the alternative clustering since they are already linked in the given clustering (See Figure 2-1).

**From Figure 2-1, it can be inferred that every possible *cannot-link* constraint has been created which is not an appropriate approach since not all of them can be fulfilled when the number of clusters is fixed (See Chapter 5.)**

On the other hand, it is important to emphasise that the side information –reference clustering- has been given a negative connotation. In other words, while in the given clustering the known associations between instances of the same clusters may be regarded as positive, they become the contrary when finding an alternative solution, i.e. the constraints negate that link.

Nonetheless, *cannot-link* constraints are not the only possible kind of constraints may be inferred from the given clustering. **Must-link** constraints indicate the pairs of instances that need to belong to the same cluster. *Must-link* constraints negate the lack of link between two instances in the given clustering. Let $x_i$, $x_j$ where $i \neq j$ be two instances in a dataset. If, in the given clustering, both instances belong to clusters $c_k, c_l$ respectively where $k \neq l$, a *must-link* constraint is created showing the aim $x_i$, $x_j$ end up belonging to the same cluster in the alternative solution. Although this approach looks plausible, issues can arise. Chapter 5 analyses them deeply.

Constrained Orthogonal Average Link Algorithm –COALA- is an instance-level constraint approach which ensures the dissimilarity between the given and the alternative clusterings by including *cannot-link* constraints in their algorithm [1]. However, dissimilarity of the alternative clustering is not the only aim of this method. Like the previous presented approaches, it also intends to discover a high-quality alternative solution. This is achieved by imposing a lowest quality threshold the alternative clustering must have. Thus, this approach also comprises a trade-off between two competitive objectives: dissimilarity and quality.

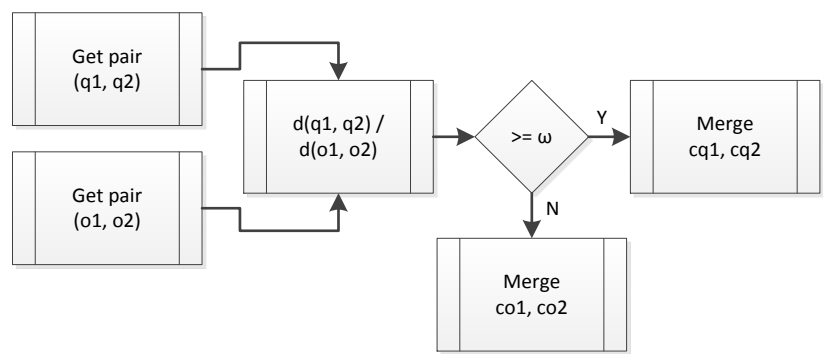

**Figure 2-2: *COALA***

*COALA* is a hierarchical agglomerative algorithm which starts its execution by considering every data object as an individual cluster –only one member. Then, an iterative process merge –agglomerates- the clusters. Each iteration of this process is depicted in this figure. Pair *(q1, q2)* corresponds to the pair of clusters that has the minimum distance between them. Pair *(o1, o2)* refers to the pair of clusters that has the minimum distance between them and also satisfies the constraints created in a pre-processing stage. A ratio is computed and compared against a pre-specified threshold ω. If this ratio is greater or equal than ω, the clusters containing instances *q1* and *q2* are merged. On the contrary, if the ratio is less than ω, clusters containing instances *o1* and *o2* are merged.

In *COALA*, a set $L$ of *cannot-link* constraints is generated previous the clustering process as such. This set contains every possible *cannot-link constraint*, that is, it contains the nodes that cannot be linked –cannot belong to the same cluster- in the alternative clustering since they are already linked in the given clustering. Afterwards, a customised hierarchical agglomerative clustering process is carried out. This algorithm is represented in Figure 2-2.

As a hierarchical agglomerative algorithm, it starts its execution by considering every data object as an individual cluster, that is, it has only one member. The notion of dissimilarity in *COALA* corresponds to the distance between clusters. *Average-linkage*[7] is the chosen technique to compute such distance. Then, within an iterative process, pairs *(q1, q2)* and *(o1, o2)* are computed. The first refers to the pair of clusters that has the minimum distance between them. The second corresponds to the pair of clusters that has the minimum distance between them and also satisfies the constraints in set $L$. A ratio between these two distances is computed and compared against the quality threshold ω. If the result is greater or equal, the algorithm merges the first pair of clusters –ensuring quality. Otherwise, it merges the second –ensuring dissimilarity.

Besides to this algorithm, [1] propose: *COALACat,* which is a variant of the original *COALA* algorithm that can be used with categorical datasets; and, a new metric to measure both dissimilarity and quality through a unique score.

It is important to mention that since *COALA* has been built upon a hierarchical clustering method, it suffers from the same limitations of that approach. For example, it cannot handle overlapping clusters and its quadratic running time might be a concern for large datasets.

Alternative Distance Function Transformation –*ADFT*- is another instance-level constraint-based approach which deals with the side information in a different manner. Instead of negating the links, in the case of instances of the same cluster, this approach does create a set of *must-link* constraints. Likewise, instead of negating the lack of links between instances that belong to different clusters, it also creates a set of *cannot-link constraints.* Hence, it is *actually characterising* the provided information through instance-level constraints.

The characterisation process is conceived as a transformation from dataset $X$ into clustering $A$. From the sets of constraints, a distance function $D$ is learnt. Finally, an alternative transformation function $D'$ is computed which converts the original dataset from which an alternative clustering $A'$ is found through any algorithm.

Obviously, this is not an exhaustive list and it only focuses on two perspectives. What is important to rescue is that the presented alternative clustering approaches include a **trade-off between two competitive objectives: quality and dissimilarity**. On the other hand, they are not suitable to find alternative clusterings within relational data since they do not take into account **structural characteristics of networks**.

---

[7] Average-linkage computes the average distance of all pairwise objects between clusters.

## 2.2. Community finding within networks

Community finding tries to discover network latent structures called communities for which there is not a universally accepted definition. Nonetheless, most work on the field consider communities as network structures whose nodes are tightly related within them but loosely related with nodes of other clusters, i.e. there are a noticeable greater number of edges between nodes within a community than the number of edges between nodes of different communities. This conception of communities becomes relevant to our work because communities can be regarded as a special case of blockmodelling where the *image matrix* is a diagonal matrix.

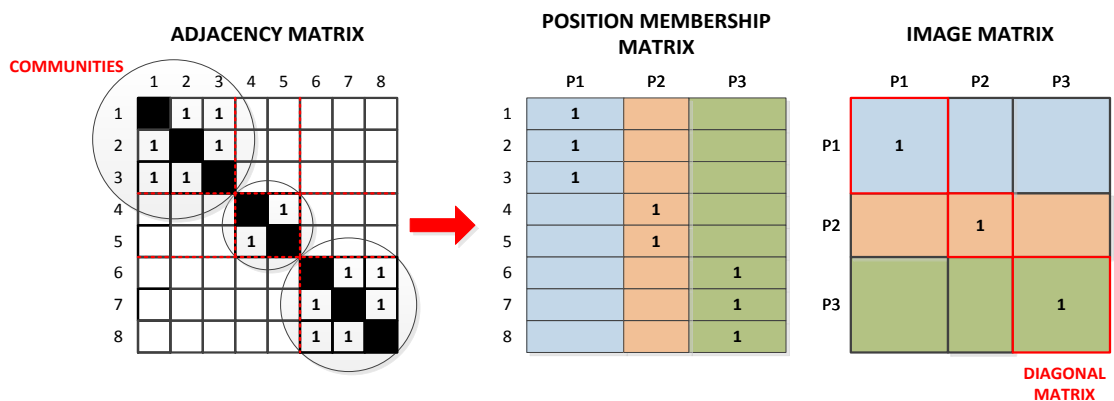

Figure 2-3: Communities and Blockmodelling

When communities are considered as structures with densely interconnected nodes within them but with sparse connections between nodes of different communities, community finding is a sub-problem within the blockmodelling field. In this figure, the adjacency matrix has three communities of vertices. Recall that the image matrix shows the relationships between positions –communities- through blocks. Thus, the image matrix is a diagonal matrix since there is no interrelations between positions in a community finding scheme.

Figure 2-3 shows an adjacency matrix which could be partitioned into three communities. In section 1.1, it was mentioned the role *image matrix* has within a blockmodel representation. Recall that it shows the relationships between positions –in this case: communities- through blocks. Conceptually, a community does not have relations with other communities –which actually does not happen. Thus, the *image matrix* in community finding is a diagonal matrix.

By far, communities have been the most studied network structure. It has been approached from different perspectives, however, it has not been solved satisfactorily yet [15]. We briefly explain some of them which we consider important within the field and for our work.

An information-theoretic approach known as *InfoMod* is presented in [26], which considers community detection as a problem of finding an efficient compression of the network. It is worth to notice that this approach is oriented to find communities by maximising mutual information between a model and the actual network. However, [26] mention that constraining the approach to find only communities will not necessarily maximise mutual information. In a second experiment they remove those constraints and find an alternative clustering.

Regarding optimisation methods, simulated annealing algorithms, greedy algorithms, spectral methods, extremal optimisation algorithms and so on, have showed noticeable results in the field of community detection [27]. Many of these approaches are based on the optimisation of a widely accepted metric called *modularity Q* introduced by [20]. *Modularity Q* can have values within the range [-1, 1] where 1 represents a network with a strong community structure – good quality. This metric is useful in divisive methods like [20] and [14] where a higher value of *Q* indicates a satisfactory split. It does not happen in hierarchical divisive clustering methods where it is not well determined when to stop the splitting process.

Nonetheless, [16] showed that optimising the *modularity Q* metric enforces to find clusters at a coarse level. It means that some clusters which are smaller than a certain threshold might not be discovered. Furthermore, the approach proposed in [20] shows an important shortcoming related with its complexity. It takes $O(n^3)$ time on sparse graphs where $n$ stands for the number of nodes.

Therefore, [13] considered improving approach's complexity in [20] by applying an iterative greedy local algorithm which maximise the *modularity Q* of the communities in each iteration. [13] demonstrated the computational efficiency of their approach by identifying language communities in a mobile phone network of 118 million nodes and more than one billion of links. Moreover, due also to its high quality results, it has become widely used [27].

[19] proposed a different approach based on modularity but using spectral principles of the network. This new approach takes $O(n^2 \log(n))$ which is considerably better. This method and many other linear approaches based on spectral clustering such as Principal Component Analysis –*PCA*-, Independent Component Analysis –*ICA*- and eigenvalue decomposition allow the existence of negative values in the resulting matrices after the factorisation process.

[22] introduced another information-theoretic algorithm based on random walks called *InfoMap* also known as *Map equation* method. This is a remarkable work within community finding field since it outperforms by far many of the proposed algorithms [17, 21]. Moreover, its rationale and its reduced time and space complexities make it a very attractive method. This approach tries to minimise the description length –*MDL*[8]- by using firstly a greedy search algorithm and then, a simulated annealing approach.

Although community finding approaches now deal with relational datasets, **they are limited to only one possible network structure, the so-called communities**. Moreover, none of these approaches tries to find an alternative community structure.

---

[8] *Minimum description length* is a principle in which the best hypothesis for a given set of data is the one that leads to the best compression of the data.

## 2.3. Blockmodelling

Within blockmodelling and Social Network Analysis contexts, similarity between nodes is determined by the so-called *structural equivalence*. Structural equivalence considers that nodes within a network are structurally equivalent, that is, belong to the same position, if they have exactly the same set of incoming and outgoing relations (See Figure 2-4). However, most real datasets do not contain nodes that are structurally equivalent in the strict sense of the definition. Therefore, the basic objective is to divide the network into positions, each of which contains nodes that are approximately structurally equivalent. For example, when comparing two pairs of nodes, the pair that is more approximately structural equivalent will be the one that has more common relations.

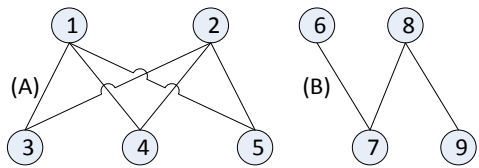

**Figure 2-4: Structural Equivalence**

(A) Nodes 1, 2, and nodes 3, 4, 5 are two clusters of structurally equivalent nodes. For example, nodes 1 and 2 can represent two teachers and nodes 3, 4 and 5 can represent their students. In this case, the teachers have the same group of students and this group of students is taught by the same two teachers. Therefore, both teachers and students are two clusters of structurally equivalent nodes. (B) None of the nodes are structurally equivalent. For instance, Twitter users 6 and 8 follow user 7, but user 8 also follows user 9. Thus, these users are not structurally equivalent.

[28] proposed a novel framework called *FactorBlock* to discover blockmodels that are present in noisy and sparse networks. [28] show that a repeated number of experiments using the algorithms introduced by [29] yield incorrect results as the factorisation algorithms penalise equally the edges and non-edges[9] -this is especially noticeable in sparse datasets. Therefore, the introduced approach includes a weighting scheme in the objective function. Likewise, noisiness is dealt by including more constraints into the function.

The problem of evolving blockmodels has also been studied, yet unlike static blockmodels, the rules to define the structural equivalence within dynamic blockmodels are not well defined. Also, within dynamic graphs is not clear what the best blockmodel is. There is a risk of representing each change with a blockmodel –overfitting- or only one block model for the whole evolutionary process –inaccuracy. Therefore, in [30], two definitions of evolving structural equivalence and an information theoretic approach to determine when a new blockmodel is necessary are proposed.

In [28], blockmodelling is related with Non-negative matrix factorisation –*NMF*-, an essential concept within our work (See section 3.1.1). In fact, [28] shows how the decomposition of an adjacency matrix could lead to the blockmodel representation. Our methodology (See Chapter 3) is based on this effective clustering process.

---

[9] Adjacency matrices of sparse datasets have many blank cells which correspond to non-edges.

Nonetheless, **any of the surveyed blockmodelling approaches deals with the problem of discovering an alternative blockmodel representation**.

## 2.4. Conclusions

Our survey has presented different approaches to solve problems related to alternative blockmodelling. From these methods, some valuable conclusions can be stated:

- Quality and dissimilarity between reference and alternative clusterings are opposite objectives so that a trade-off parameter allows controlling their effect on the whole objective.
- Community finding is a special case of blockmodelling where the image matrix is a diagonal matrix.
- Blockmodelling techniques identify positions –clusters- by evaluating structural characteristics of the nodes within a network. The most important equivalence metric in this field is called *structural equivalence*.

# Chapter 3

# Methodology

Based on the proposed problem (See section 1.2.1,) our approach must find an alternative blockmodel representation of a network. This representation is a **summary** of the adjacency matrix of the network through two sub-matrices: *position membership matrix* and *image matrix.* Then, decomposition of the adjacency matrix into factor matrices of **lower rank** can be thought as a plausible approach to find a summary of the adjacency matrix. For example, *PCA* is a widely used method for **dimensionality reduction** by performing either eigenvalue decomposition[10] of a data covariance matrix[11] or Singular Value Decomposition –*SVD*- of a data matrix.

Nonetheless, *PCA* or *SVD* do not constrain the range of the values within the factor matrices. In other words, the factor matrices might have negative values. Let $X \in \mathbb{R}^{p \, x \, n}$ be the adjacency matrix of a network of $n$ vertices. Then,

$$X \approx U \Sigma V^*$$

is the singular value decomposition of $X$ into two sub-matrices $U \in \mathbb{R}^{p*k}$ and $V \in \mathbb{R}^{n*k}$ where $k$ is chosen to hold: $k * (p + n) \ll p * n$.

Although a summary of the adjacency matrix is achieved through the dimensionality reduction, the presence of negative values in any of the factor sub-matrices makes interpretation harder within a clustering context. Then, a positive matrix factorisation process would accomplish this requirement. But first,

- Can we consider matrix factorisation of the adjacency matrix a clustering process?
- How is matrix factorisation related with blockmodelling?
- Does matrix factorisation identify structural characteristics of the nodes in order to determine similarity/dissimilarity between them?

The next section presents fundamental concepts to our methodology which can answer these questions.

---

[10] Eigenvalue decomposition is the factorization of a matrix into eigenvalues and eigenvector matrices.
[11] Covariance matrix $C \in \mathbb{R}^{n \, x \, n}$ has each cell $c_{ij} = cov(i^{th} \; vector, j^{th} \; vector)$ where $i^{th}$ and $j^{th}$ are column vectors within dataset $D \in \mathbb{R}^{m \, X \, n}$.

## 3.1. Fundamental concepts

### 3.1.1. Non-negative matrix factorisation NMF

Non-negative matrix factorisation –*NMF*- decomposes a matrix into factor sub-matrices with non-negativity constraints. Therefore, *NMF* is an approximation of the form

$$X \approx YZ^T$$

where $X \in \mathbb{R}_+^{p*n}$, $Y \in \mathbb{R}_+^{p*k}$ and $Z \in \mathbb{R}_+^{n*k}$

$X$ is a matrix with non-negative values that could be mapped to the adjacency matrix of a network. Then, the factorisation process becomes an optimisation problem which minimises the error of the approximation $X \approx YZ^T$.

In [10], the authors present *NMF* with its non-negativity constraints. In fact, they demonstrate the importance *NMF* constraints have as they guide to a parts-based representation of a whole. By restraining to non-negative values, only additive combinations would be present. Hence, *NMF* is indeed more appropriate when it is compared against linear methods such as *PCA* and *SVD*. However, the question arises again:

*Can we consider NMF of the adjacency matrix a clustering process?*

Experiments carried out by [31] and [32] showed the effectiveness of *NMF* in clustering and pattern recognition. Moreover, [9] demonstrated that *NMF* is related to *k-means* clustering and graph partitioning methods. Therefore, *NMF* is an appropriate clustering tool of relational data which has the added value of being interpretable.

Nevertheless, the approximation $X \approx YZ^T$ does not resemble the blockmodel configuration.

Thus:

*How is NMF related with blockmodelling?*

[29] and [33] propose non-negative **matrix tri-factorisation** –*NMTF*- as an option to deal with the blockmodel problem. *NMTF* is the decomposition of a matrix into three factor sub-matrices. Its general form is shown below:

$$X \approx FMG^T$$

where $X \in \mathbb{R}_+^{p \times n}$, $F \in \mathbb{R}_+^{p \times k}$, $M \in \mathbb{R}_+^{k \times l}$ and $G \in \mathbb{R}_+^{n \times l}$. In this case, $k \neq l$ but it is more common that $k = l$ [9].

However, this general form does not map to the blockmodel configuration. In order to overcome this issue, a useful consideration is that $X = X^T = A$ since $X$ represents the adjacency matrix and thus, it is symmetric[12]. As a consequence $F = G = C$. Then, the symmetric non-negative matrix tri-factorisation –*SNMTF*- form follows:

$$A \approx CMC^T$$

**$A$ represents the adjacency matrix, $C$ the *position membership matrix* and $M$ the image matrix**. Hence, *SNMTF* maps nicely to the blockmodel configuration (See Figure 3-1).

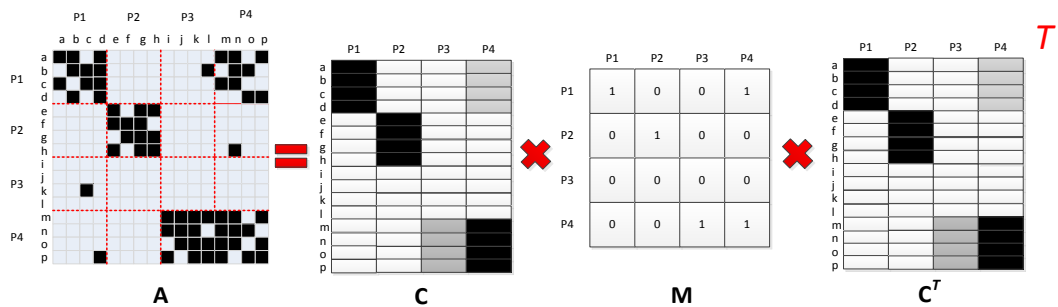

**Figure 3-1: Symmetric Non-negative Matrix Tri-Factorisation SNMTF**

This figure shows a graphical representation of the symmetric nonnegative matrix tri-factorization approximation. An adjacency matrix is decomposed into three sub-matrices.

*Does SNMTF identify structural characteristics of the nodes in order to determine similarity/dissimilarity between them?*

As stated in section 2.3, *structural equivalence* is the most used criteria in Social Network Analysis when evaluating common structural characteristics of nodes within a network. Recall that *structural equivalence* considers that two nodes within a network are structurally equivalent if they have exactly the same set of incoming and outgoing relations. This conception of equivalence between nodes imposes a challenge to conventional clustering algorithms on networks. For instance, let's consider a two-mode network of students and subjects taken by these students. A graph of this example is shown in Figure 3-2.

In this example, traditional clustering algorithms could not be able to discover clusters of only students or clusters of only subjects. The reason is because there are no explicit relations between students neither between subjects. However, *S1, S2* and *S3* are structurally equivalent since they have the same set of taken subjects –incoming and outgoing relations.

The good news is that *NMF* is able to deal with *structural equivalence* as it is demonstrated by [34] on several real network datasets.

Therefore, **NMF and its special case *SNMTF* are the appropriate tools in order to face the proposed problem** (See section 1.2.1.) **These tools have the competence for clustering relational data. They also fit nicely with the blockmodelling configuration; and, they can deal**

---

[12] **WARNING: We are considering in our work only undirected networks.**

**with the challenge of equivalence in networks. Furthermore, their ability to handle structural equivalence allows them to discover network structures other than communities.**

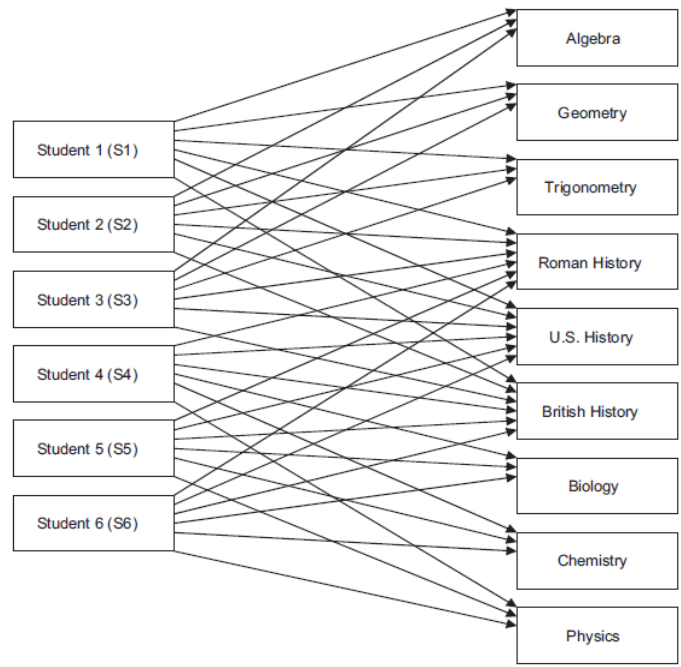

Figure 3-2: Two-mode graph of students and subjects [34]

Finally, it is worth to mention *NMF* offers advantages as a clustering process:

- *NMF* can be used in different types of networks such as non-directional, e.g. a social network like Facebook®; and, directional, e.g. email network.
- *NMF* is able to manage soft and hard[13] clustering. In the case of soft clustering, due to the fact that *NMF* is an approximation of an adjacency matrix through the multiplication of three sub-matrices, the real values within the position membership matrix could be normalised as probabilities of membership of each cluster. On the contrary, to achieve hard clustering, it is only necessary to include orthogonality constraints between the columns of the position membership matrix. Afterwards, *NMF* is applied with no variation.

---

[13] In soft clustering, nodes within a network can belong to more than one cluster. Hard clustering allows nodes to belong to only one cluster.

### 3.1.2. Solving SNMTF

In [11], *NMF* is regarded as an optimization problem of either the conventional least squares error or the Kullback-Leibler divergence[14] -$D_{KL}$-. Similarly, [29] consider *NMTF* as a least squares error problem. In both cases, the problem is solved by minimising the corresponding loss function –least squares error or $D_{KL}$

Since our methodology is based on *SNMTF*, we present the initial objective function that minimises the least squares error between the adjacency matrix and its approximation -the multiplication of three sub-matrices.

$$\min_{C\geq 0, M\geq 0} \|A - CMC^T\|_F^2 \qquad (1)$$

where $\|\cdot\|_F^2$ is the square of the Frobenius norm[15], $A \in \mathbb{R}_+^{n \times n}, C \in \mathbb{R}_+^{n \times k}$ and $M \in \mathbb{R}_+^{k \times k}$.

[11] observed that the optimisation problem is convex if one of the two matrices $C$ or $M$ is fixed; otherwise, when both $C$ and $M$ are variable, the problem becomes not convex. Thus, **SNMTF is a not convex optimisation problem for which only a local optimum solution can be found.** [11] also presented a relative efficient iterative algorithm called *multiplicative update rules* to estimate a local optimum. These rules can be thought as a rescaled *gradient descent*[16] algorithm. They guarantee the accomplishment of the non-negativity constraints since no subtraction is present –which does not happen in the gradient descent algorithm where variable values might change by subtraction.

In our methodology, multiplicative update rules are obtained by restating the objective function (1). Since this optimisation problem encompasses inequality constraints –non negativity constraints-, Karush-Kuhn-Tucker –*KKT*- multipliers are introduced as new variables into (1). Hence, the initial *SNMTF* objective function becomes:

$$F(C, M, \mu_1, \mu_2) = \|A - CMC^T\|_F^2 - tr(\mu_1 C^T) - tr(\mu_2 M^T) \qquad (2)$$

where $tr(\cdot)$ is the trace[17] of a matrix and $\mu_1$ and $\mu_2$ are the *KKT* multipliers.

The introduction of *KKT* multipliers is a strategy to find a local optimum of a function subject to inequality constraints. *KKT* multipliers are variables within a system of equations corresponding to conditions –*KKT* conditions- for a solution to be optimal. An analytical solution to this system of equations is not trivial, thus the multiplicative update rules provide a mechanism to find an approximation.

---

[14] Kullback-Leibler divergence $D_{KL}(P||Q)$ measures the dissimilarity between two probability distributions $P$ and $Q$: $D_{KL}(P||Q) = \sum_i ln\left(\frac{P(i)}{Q(i)}\right) P(i)$

[15] Frobenius norm is also called Euclidean distance norm and it is obtained as follows: $\|A\|_F = \sqrt{\sum_{i=1}^{n} \sum_{j=1}^{m} |a_{ij}|^2}$.

[16] Gradient descent is an optimisation algorithm which finds a local minimum. First of all, the gradient of the objective function at the current point is obtained –if it is possible, otherwise, it is estimated. Then, the independent variables changes in the direction of the negative gradient.

[17] Trace of a matrix $A$: $tr(A) = \sum_i a_{ii}$

For convenience, in Equation (2), $\|\cdot\|_F^2$ can be expressed by using $tr(\cdot)$ as:

$$F(C, M, \mu_1, \mu_2) = tr(A^T A) - 2tr(A^T CMC^T) + tr(CM^T C^T CMC^T) - tr(\mu_1 C^T) - tr(\mu_2 M^T) \quad (3)$$

The partial derivatives of function (3) with respect to $C$ and $M$ are:

$$\frac{\partial F}{\partial \boldsymbol{C}} = 2(CM^T C^T CM + CMC^T CM^T) - 2(A^T CM + ACM^T) - \mu_1 \quad (4)$$

$$\frac{\partial F}{\partial \boldsymbol{M}} = 2(C^T CMC^T C) - 2(C^T AC) - \mu_2 \quad (5)$$

The partial derivatives of function (3) with respect to the *KKT* multipliers are:

$$\frac{\partial F}{\partial \boldsymbol{\mu_1}} = -C; \quad \frac{\partial F}{\partial \boldsymbol{\mu_2}} = -M$$

When $\frac{\partial F}{\partial \mu_1}$ and $\frac{\partial F}{\partial \mu_2}$ are equalled to zero, the resulting equations refer to the boundaries of the inequality constraint functions. Thus, these derivatives are not regarded as part of the system of equations from which the multiplicative update rules are defined.

The multiplicative update rules for solving equation (3) are shown below:

$$C_{ij} \leftarrow C_{ij} \left[ \frac{A^T CM + ACM^T}{CM^T C^T CM + CMC^T CM^T} \right]_{ij} \quad (6)$$

$$M_{ij} \leftarrow M_{ij} \left( \frac{C^T AC}{C^T CMC^T C} \right)_{ij} \quad (7)$$

Equations (6) and (7) are computed iteratively until a **stop criterion** is reached. Normally, this criterion is the extent of change of the value of the objective function from the previous to the current iteration. If the change is less than a threshold, e.g. $< 10^{-5}$, it is reasonable to think that the algorithm has converged; then, the algorithm can stop.

However, *NMF* is an **initialisation-sensitive** algorithm. Its efficiency depends mostly on which point in the search space the matrices $C$ and $M$ start off. Therefore, in our methodology, the algorithms are run many times with random starting points for those matrices.

Up to this point, no alternative-blockmodelling-related strategy has been included in the equations. In this chapter, as part of our methodology we derive two different approaches which append new objectives to the main objective function (1): *(a) inclusion of cannot-link constraints*, and *(b) dissimilarity between image matrices*.

But before diving into our approaches, let's discuss two possible normalisation strategies.

### 3.1.3. Normalisation vs. Iterative Lagrangian Solution

Random initialisation of the matrices $C$ and $M$ at the beginning of the algorithm injects uncertainty on finding a local minimum for the objective function. But, this is not the only reason why many different possible solutions could be found after running iteratively the multiplicative update rules. As noticed by [35], in the simpler *NMF* configuration

$$F = CH^T$$

where $F \in \mathbb{R}_+^{n \times m}, C \in \mathbb{R}_+^{n \times k}$ and $H \in \mathbb{R}_+^{m \times k}$.

there exist a large number of matrices $(A, B)$ such that $AB^T = I, CA \geq 0, HB \geq 0$. Hence, $(CA, HB)$ is a possible solution since it has the same function value $F$. **Normalisation of the factor matrices is an action to remove part of the uncertainty**.

Regarding the *SNMTF* configuration

$$A = CMC^T$$

where $A \in \mathbb{R}_+^{n \times m}, C \in \mathbb{R}_+^{n \times k}$ and $M \in \mathbb{R}_+^{k \times k}$.

Intuitively, the *position membership* matrix $C$ represents the probability of the nodes to belong to each cluster $c_l, 1 \leq l \leq k$. Therefore, matrix $C$ is a *right stochastic*[18] matrix where

$$\sum_{l=1}^{k} C_{il} = 1; \; \forall i, 1 \leq i \leq n \qquad (8)$$

Therefore, because of removing uncertainty and following the intuition of stochasticity, **factor matrices must be normalised**. [12] propose two methods besides to conventional normalisation since the authors demonstrate that **conventional normalisation suffers from leading to potential poor local optimum and slow convergence rate**.

From those two methods, our methodology uses the *relaxation* version of the stochasticity constraints. Then, the objective function (1) becomes

---

[18] There are also *left* and *vectorised stochastic* matrices. In a *left stochastic* matrix, the values of the column $i$ sum up to 1 for all $i$. In a *vectorised stochastic* matrix, all values sum up to 1.

$$\min \|A - CMC^T\|_F^2$$

$$\text{subject to} \quad \sum_{l=1}^{k} C_{il} = 1; \forall i, 1 \leq i \leq n \tag{9}$$

$$C \geq 0, M \geq 0$$

In order to deal with the inequality constraints and the new set of equality constraints, *KKT* and *Lagrange* multipliers are introduced into the objective function (9). Then, this function, with its constraints, is equivalent to

$$F(C, M, \{\lambda_i\}_{i=1}^n, \mu_1, \mu_2) = \|A - CMC^T\|_F^2 + \sum_{i}^{n} \lambda_i \left( \sum_{l=1}^{k} C_{il} - 1 \right) - tr(\mu_1 C^T) - tr(\mu_2 M^T) \tag{10}$$

where $\{\lambda_i\}_{i=1}^n$ are the *Lagrange* multipliers.

The partial derivative of function (10) with respect to $M$ remains the same (See equation (5)), whereas there is a slight variation in the partial derivative of the function with respect to $C$ (See equation (4)) as follows:

$$\frac{\partial F}{\partial C} = 2(CM^T C^T CM + CMC^T CM^T) - 2(A^T CM + ACM^T) - \Lambda - \mu_1 \tag{11}$$

where $\Lambda = [\lambda_{ij}]$ is the matrix of *Lagrange* multipliers.

Following the *iterative Lagrangian* solution proposed in [12], variables of function $F$ can be solved as follows:

---

a. A preliminary update rule

$$C'_{ij} \leftarrow C_{ij} \frac{\nabla^-_{ij} + \lambda_i}{\nabla^+_{ij}}$$

where
$$\nabla^+ = (CM^T C^T CM + CMC^T CM^T); \quad \nabla^- = A^T CM + ACM^T$$

b. This preliminary update rule replaces the corresponding term in the equality constraint $\sum_{b=1}^{k} C_{ib} = 1$

$$\sum_{b=1}^{k} C_{ib} \frac{\nabla^-_{ib}}{\nabla^+_{ib}} + \lambda_i \sum_{b=1}^{k} \frac{C_{ib}}{\nabla^+_{ib}} = 1$$

c. Solving for $\lambda_i$

$$\lambda_i = \frac{1 - \sum_{b=1}^{k} C_{ib} \frac{\nabla^-_{ib}}{\nabla^+_{ib}}}{\sum_{b=1}^{k} \frac{C_{ib}}{\nabla^+_{ib}}}$$

---

> d. In the preliminary update rule, $\lambda_i$ is replaced
>
> $$C_{ij} \leftarrow C_{ij} \frac{\nabla^-_{ij} G_{ij} + 1}{\nabla^+_{ij} G_{ij} + H_{ij}}$$
>
> where (12)
>
> $$G_{ij} = \sum_b \frac{C_{ib}}{\nabla^+_{ib}}; \quad H_{ij} = \sum_b C_{ib} \frac{\nabla^-_{ib}}{\nabla^+_{ib}}$$

Table 1: Iterative Lagrangian solution [12]

Equation (12) is the new multiplicative update rule for $C$ using the *iterative Lagrangian* approach. It is worth to mention that the "moving term" trick [36] was applied in step (d).

The multiplicative update rule for $M$ remains the same (See equation (7)).

### 3.2. Outline of our Methodology

Once we have presented both the suitability of *NMF* –particularly *SNMTF*- to solve the proposed problem (See section 1.2.1) and the algorithm –multiplicative update rules- to optimise it, we outline our algorithm at a high level,:

> **Step 1:** Uniform random initialisation of matrices $M$ and $C$ with nonnegative values.
>
> **Step 2:** Update $M$
>
> **Step 3:** Update $C$
>
> Iterative multiplicative update rules DEPEND ON THE CHOSEN APPROACH. Our methodology comprises two approaches: *(a) inclusion of cannot-link constraints* (See section 3.3), and *(b) dissimilarity between image matrices* (See section 3.4)
>
> **Step 4:** Verify the stop criterion
>
> If the change of the objective function value –from previous to current iteration- is less than a threshold, e.g. $< 10^{-5}$, the loop ends; otherwise a new iteration starts from Step 2.

Table 2: Outline of our algorithm

In each of our two approaches, the objective function stated in (9) is expanded by adding different objectives which encode the alternative blockmodelling purpose.

Finally, *evaluation of the alternative blockmodelling solution* is carried out in order to show the achievement of the quality and dissimilarity requirements.

## 3.3. Approach 1: Inclusion of *cannot-link* constraints

As mentioned in section 2.1.2, instance-level constraints encode side-information as pairs of instances –nodes in relational data- in the form of *cannot-link* and *must-link* constraints. With respect to our problem, side-information refers to the given or reference clustering. Thus, instance-level constraints translate the information provided by the reference clustering into *cannot-link* and *must-link* constraints.

Nonetheless, in this approach, we decided to work only with *cannot-link* constraints (See Chapter 5 for details.) Moreover, we regard the reference clustering as **negative information** about the sought alternative clustering. That is, this approach is similar to the one proposed by [1] where a pair of instances that belong to the same cluster in the reference clustering becomes a *cannot-link* constraint. In this way, the clustering process is guided to discover a dissimilar clustering with respect to the given one.

Therefore, *cannot-link* constraints can be included into the objective function (9) in such a way that violation of these constraints is penalised, i.e.

$$\min_{C \geq 0, M \geq 0} \|A - CMC^T\|_F^2 + tr(\beta C^T \Theta C)$$

$$\text{subject to} \quad \sum_{l=1}^{k} C_{il} = 1; \; \forall i, 1 \leq i \leq n \qquad (13)$$

$$C \geq 0, M \geq 0$$

where $\Theta$ corresponds to the *cannot-link* constraints matrix and $\beta$ is a trade-off parameter between the quality term $\|A - CMC^T\|_F^2$ and the dissimilarity term $tr(\beta C^T \Theta C)$.

The effect of the new term $tr(C^T \Theta C)$ within the objective function (13) is shown in Example 1:

**Example 1:**

$$\Theta = \begin{bmatrix} 0 & \theta & 0 \\ \theta & 0 & 0 \\ 0 & 0 & 0 \end{bmatrix}, C^t = \begin{bmatrix} 1 & 0 \\ 1 & 0 \\ 0 & 1 \end{bmatrix}$$

This example shows a *cannot-link* constraints matrix $\Theta$ that is built based on the reference clustering. This matrix states the aim of break the link between nodes 1 and 2 since a value $\theta > 0$ was placed in positions (1, 2) and (2, 1). It also shows a *position membership* matrix obtained at time $t$ through the multiplicative update rules. Thus, if at this time $t$ the objective function (13) is evaluated, the expression

$$tr\left(C^{t^T} \Theta C^t\right) = tr\left(\begin{bmatrix} 1 & 1 & 0 \\ 0 & 0 & 1 \end{bmatrix} \begin{bmatrix} 0 & \theta & 0 \\ \theta & 0 & 0 \\ 0 & 0 & 0 \end{bmatrix} \begin{bmatrix} 1 & 0 \\ 1 & 0 \\ 0 & 1 \end{bmatrix}\right) = tr\left(\begin{bmatrix} 2\theta & 0 \\ 0 & 0 \end{bmatrix}\right) = 2\theta$$

> Whereas, if at time $t+1$
>
> $$C^{t+1} = \begin{bmatrix} 1 & 0 \\ 0 & 1 \\ 0 & 1 \end{bmatrix}$$
>
> Then, the expression
>
> $$tr\left(C^{t+1^T}\Theta C^{t+1}\right) = tr\left(\begin{bmatrix} 1 & 0 & 0 \\ 0 & 1 & 1 \end{bmatrix}\begin{bmatrix} 0 & \theta & 0 \\ \theta & 0 & 0 \\ 0 & 0 & 0 \end{bmatrix}\begin{bmatrix} 1 & 0 \\ 0 & 1 \\ 0 & 1 \end{bmatrix}\right) = tr\left(\begin{bmatrix} 0 & \theta \\ \theta & 0 \end{bmatrix}\right) = 0$$
>
> Hence, the effect of $tr(C^T\Theta C)$ is indeed the expected, i.e. it penalises the violation of the *cannot-link* constraint by incrementing the value of the objective function at time $t$ by $2\theta$.

This approach presents two opposite objectives that represent quality and dissimilarity respectively. Hence, a trade-off parameter $\beta$ is needed to balance the effect of the dissimilarity term on the whole objective.

After the inclusion of the new term, the partial derivative (11) shifts to

$$\frac{\partial F}{\partial C} = 2(CM^TC^TCM + CMC^TCM^T) - 2(A^TCM + ACM^T) - \Lambda - \mu_1 + 2\Theta C \qquad (14)$$

Whereas, the partial derivative (5) remains the same.

Consequently, the multiplicative update rules for matrices $C$ and $M$ are

$$C_{ij} \leftarrow C_{ij}\frac{\nabla^-_{ij}G_{ij} + 1}{\nabla^+_{ij}G_{ij} + H_{ij}}$$

where

$$G_{ij} = \sum_b \frac{C_{ib}}{\nabla^+_{ib}}; \quad H_{ij} = \sum_b C_{ib}\frac{\nabla^-_{ib}}{\nabla^+_{ib}}$$

$$\nabla^+ = (CM^TC^TCM + CMC^TCM^T + \boldsymbol{\beta\Theta C}); \quad \nabla^- = A^TCM + ACM^T$$

(15)

$$M_{ij} \leftarrow M_{ij}\left(\frac{C^TAC}{C^TCMC^TC}\right)_{ij} \qquad (16)$$

## 3.4. Approach 2: Dissimilarity between image matrices

Quality and dissimilarity are the objective functions that compete within the alternative clustering scheme as it was shown in [1, 3, 4, 7]. Our first approach also leverages that concept. This second approach inherits this idea but within the context of blockmodelling. Thus, quality is assured by the least squares error, whereas dissimilarity is addressed by the intuition of differences between *image* matrices. Recall that *image* matrix in blockmodelling shows the relationships between positions –clusters-, that is, an image matrix shows the network structure as a sort of summary. Then, a different image matrix would correspond to a different network structure. For example, if our given clustering has a community structure represented by its *image* matrix, it would make sense to "negate" that structure through an objective function that maximises the difference between the *image* matrices.

Therefore, the term representing the dissimilarity between the given and the alternative *image* matrices is included into the base objective function (9) as follows

$$\min \|A - CMC^T\|_F^2 - \beta \|M^{(0)} - M\|_F^2$$

$$\text{subject to} \quad \sum_{l=1}^{k} C_{il} = 1; \ \forall i, 1 \leq i \leq n \quad (17)$$

$$C \geq 0, M \geq 0$$

where $M^{(0)} \in \mathbb{R}_+^{k \times k}$ corresponds to the *image* matrix of the given clustering. Again, $\beta$ is a trade-off parameter between the quality term $\|A - CMC^T\|_F^2$ and the dissimilarity term $\|M^{(0)} - M\|_F^2$.

After the inclusion of the new term, the partial derivative (5) shifts to

$$\frac{\partial F}{\partial M} = 2(C^T CMC^T C) - 2(C^T AC) - \mu_2 + 2\beta(M^{(0)} - M) \quad (18)$$

The partial derivative with respect to $C$ does not change (See equation (11)). As a consequence, the multiplicative update rules for matrices $C$ and $M$, for our second approach are

$$C_{ij} \leftarrow C_{ij} \frac{\nabla^-_{ij} G_{ij} + 1}{\nabla^+_{ij} G_{ij} + H_{ij}}$$

where

$$G_{ij} = \sum_b \frac{C_{ib}}{\nabla^+_{ib}}; \quad H_{ij} = \sum_b C_{ib} \frac{\nabla^-_{ib}}{\nabla^+_{ib}}$$

$$\nabla^+ = (CM^T C^T CM + CMC^T CM^T); \quad \nabla^- = A^T CM + ACM^T$$

(19)

$$M_{ij} \leftarrow M_{ij} \left( \frac{C^T AC + \boldsymbol{\beta M}}{C^T CMC^T C + \boldsymbol{\beta M^{(0)}}} \right)_{ij}$$

(20)

### 3.4.1. Generalised form

This second approach can be generalised to find an alternative blockmodel which is different from a set of given blockmodels. This set of reference clusterings might be constructed incrementally by using even any other approach. Hence, the generalised form of the objective function (17) is as follows

$$\min \|A - CMC^T\|_F^2 - \sum_{p=0}^{|\mathfrak{B}|} \boldsymbol{\beta_p} \|M^{(p)} - M\|_F^2$$

subject to  $\sum_{l=1}^{k} C_{il} = 1; \forall i, 1 \leq i \leq n$

$C \geq 0, M \geq 0$

(21)

where $|\mathfrak{B}|$ corresponds to the number of given blockmodels –*image* matrices.

The multiplicative update rule for matrix $M$ is

$$M_{ij} \leftarrow M_{ij} \left( \frac{C^T AC + M \sum_{p=0}^{|\mathfrak{B}|} \boldsymbol{\beta_p}}{C^T CMC^T C + \sum_{p=0}^{|\mathfrak{B}|} \boldsymbol{\beta_p} M^{(p)}} \right)_{ij}$$

(22)

The multiplicative update rule for matrix $C$ remains the same as (19).

## 3.5. Evaluation

Now, *how can we evaluate the effectiveness of our approaches?*

Since the objective function of both approaches comprise a trade-off between quality and dissimilarity, the discovered alternative blockmodel must be quantitatively evaluated from these two perspectives.

With respect to *quality*, external evaluation might be hard to be carried out because the *gold-standard* is unlikely to be known, especially when we are trying to discover a novel blockmodel representation. On the other hand, current internal evaluation measures are not well-suited to blockmodelling as a fitting model of the data. However, in [30] the concept of *MDL* is adapted to the blockmodelling problem. It regards that **a blockmodel is a suitable model if its encoding cost is low**. The total encoding cost of a blockmodel is determined by the aggregate of the costs of its positions and its blocks, and it is called *Individual Snapshot Encoding* in [30]. Then, the total encoding cost of a blockmodel $\mathfrak{B}$ is

$$C_{ind}(\mathfrak{B}) = C(n) + C(\mathcal{C}) + C(\mathcal{G}|\mathcal{C})$$

where $C(n)$ is the cost of sending the number of vertices $n$ –it is necessary for reconstruction reasons[19]-, $C(\mathcal{C})$ is the cost of encoding the positions, and $C(\mathcal{G}|\mathcal{C})$ is the cost of encoding the blocks.

We refer to [30] for details on computation of the partial costs.

In the case of dissimilarity, as stated by [37], there are plenty of metrics that might be used when comparing two clusterings. However, all of them fail to consider particular structural characteristics of blockmodels. For instance, when comparing two clusterings by computing the Normalised Mutual Information[20] –*NMI-*, this metric might consider them as similar when actually, the density of the blocks between certain positions is quite different. Therefore, [37] proposes some structure-aware distance measures from which we are using the one called *Kullback-Leibler Reconstruction Measure*. It is based on the Kullback-Leibler divergence between matrices. Therefore, this measure is asymmetric and it can be regarded as using the edge distributions in the second blockmodel to encode the edge distributions in the first. The edge reconstruction KL distance is defined as

$$d_{RKL}\left(\mathfrak{B}^{(0)}, \mathfrak{B}^{(1)}\right) = \sum_i \sum_j \hat{A}_{i,j}^{(0)} * \log\left(\frac{\hat{A}_{i,j}^{(0)}}{\hat{A}_{i,j}^{(1)}}\right) - \hat{A}_{i,j}^{(0)} + \hat{A}_{i,j}^{(1)}$$

where $\hat{A}$ is the adjacency matrix approximation, that is, $\hat{A} = CMC^T$.

---

[19] *MDL,* as an information theory concept, regards that the data is being compressed to be sent through a communication channel and reconstructed at the destination.
[20] *NMI* is a measure of mutual dependence between two random variables.

As our approaches are based on *NMF*, they are initialisation-sensitive methods, thus **many runs are needed to find out a tendency in the clustering process**. We support our evaluation process by using a tool called *Enhanced Visual Analysis for Cluster Tendency –iVAT-*, which presents those many alternative blockmodels –one for each run- in a graphical form. *iVAT* was proposed by [38] and is used to estimate the number of possible clusters prior to the clustering process as such by using the pairwise dissimilarity matrix of objects to be clustered. This tool reorders the rows and columns of this dissimilarity matrix in such a way that the least dissimilar –most similar- objects appear as a set of dark blocks along the diagonal of this matrix.

As a variation of this form of using *iVAT*, we run $n$ possible blockmodel solutions by using any of our approaches and instead of creating a dissimilarity matrix of objects; we create a dissimilarity matrix of blockmodels. This meta-clustering dissimilarity matrix is reordered by *iVAT* and shows potential alternative solutions to be exploited along its diagonal. Figure 3-3 shows an example.

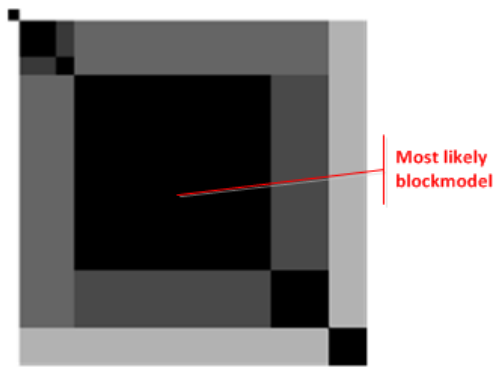

**Figure 3-3: iVAT graphic which shows some alternative blockmodels**

In this dissimilarity matrix, each column and row corresponds to a blockmodel solution obtained from any of our two approaches. The blockmodels can be compared by using any "distance" metric.

Therefore, our whole evaluation process consists of computing two quantitative metrics for the quality and dissimilarity requirements. **Low $C_{ind}$ and high $d_{RKL}$ values are sought.** Finally, *iVAT* is used as support in order to verify if the chosen alternative blockmodel is likely to be obtained. It is worth to mention that **we are not regarding that the chosen solution needs to be statistically significant within this "sampling" process of 100 runs. At the end, which really matters, is to find a dissimilar and good quality solution even if it its occurrence is small.**

# Chapter 4

# Experiments

In this section, we show experiments carried out on real and synthetic sets of relational data. **These experiments are focused on reinforcing the contributions stated in section 1.2.**

First of all, we show the importance of the proposed problem, *"Discovering an alternative blockmodel solution in a graph"*, on two real datasets: *Zachary's karate club* [39] and *Books on American politics* [19]. *Zachary's karate club* dataset is the compilation of the relations between members of a karate club in an American university where there was an internal conflict. The instructor and the president of the club had a confrontation which caused the division of the members into 2 groups –2 communities. This dataset consists of 34 members – nodes- whose relations –edges- are undirected and unweighted. On the other hand, *Books on American politics* is a dataset of books bought on-line in Amazon.com. In this real network, books are nodes and an edge between two books appears if these two books were purchased by the same buyer. There are 105 books which span the American political alignments, i.e. liberal, conservative or centrist.

Secondly, we compare the performance of our two approaches, (1) *Inclusion of cannot-link constraints* (See section 3.3), and (2) *Dissimilarity between image matrices* (See section 3.4); on a synthetic dataset which has 50 nodes and whose edges are undirected and weighted.

Thirdly, through the application of our second approach on the same synthetic dataset, we give some insights into the difficulties of this approach and end up with some interesting conclusions.

Finally, all the experiments were run on an Intel Core i7 2.00GHz PC with 8GB of memory and running Windows 7.

## 4.1. The novel Alternative Blockmodelling problem

*Zachary's karate club* has been adopted as a benchmark in community finding field since it offers a neat community network structure. It comprises two well defined communities which align with either the instructor or the president of the club. Figure 4-1 shows this real network where communities are depicted with different colours.

On the other hand, Figure 4-2 shows the blockmodel representation of the network. Again, this representation corresponds to the actual division of the network. Note that the dense blocks are aligned through the diagonal of the matrix.

Therefore, by using any of our two approaches, the goal would be to discover an alternative blockmodel representation of *Zachary's karate club* regarding the actual blockmodel representation in Figure 4-2 as the reference clustering. **For convenience, we are using the second approach on this section.**

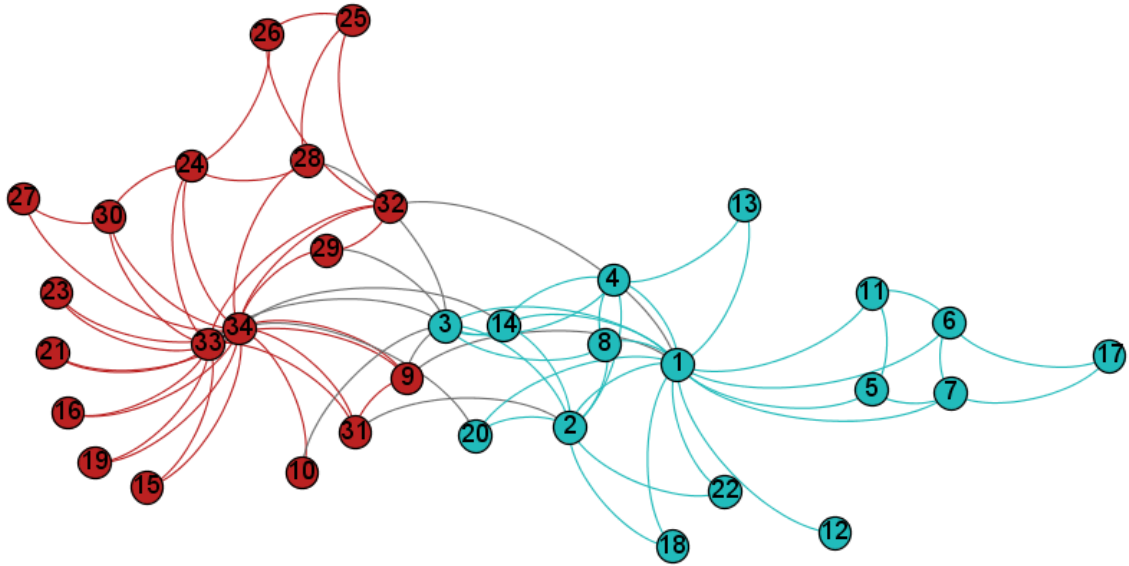

**Figure 4-1: Zachary's karate club.**

This network shows two well defined communities. Recall that a community is a cluster of nodes where they are densely connected, but less connected with nodes of other communities.

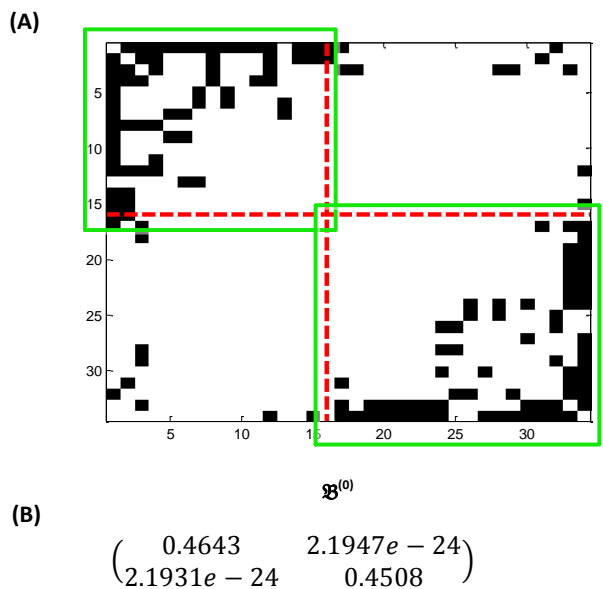

(A)

(B)

$$\begin{pmatrix} 0.4643 & 2.1947e-24 \\ 2.1931e-24 & 0.4508 \end{pmatrix}$$

$\mathfrak{B}^{(0)}$

**Figure 4-2: Zachary's karate club blockmodel representation of the actual division.**

(A) This blockmodel representation of *Zachary's karate club* shows the actual division of the network into two well defined communities. Note that the two dense blocks are aligned through the diagonal of the matrix. These two dense blocks not only represent the communities within the network but also the lack of relation of these clusters with other clusters. At the beginning of the experimental stage, this blockmodel is regarded as the given blockmodel $\mathfrak{B}^{(0)}$.

(B) Image matrix corresponding to this blockmodel representation.

Hence, the initial image matrix $M^{(0)}$ is $\begin{pmatrix} 0.4643 & 2.1947e-24 \\ 2.1931e-24 & 0.4508 \end{pmatrix}$. Approach 2 balances the effect of the dissimilarity term on the whole objective function by using a trade-off parameter $\boldsymbol{\beta}$. Therefore, the algorithm is run with different values for this parameter. Figure 4-3 shows alternative blockmodel representations of the *Zachary's karate club* network with respect to the given image matrix $M^{(0)}$ for different values of the $\boldsymbol{\beta}$ parameter.

(A) β = 1

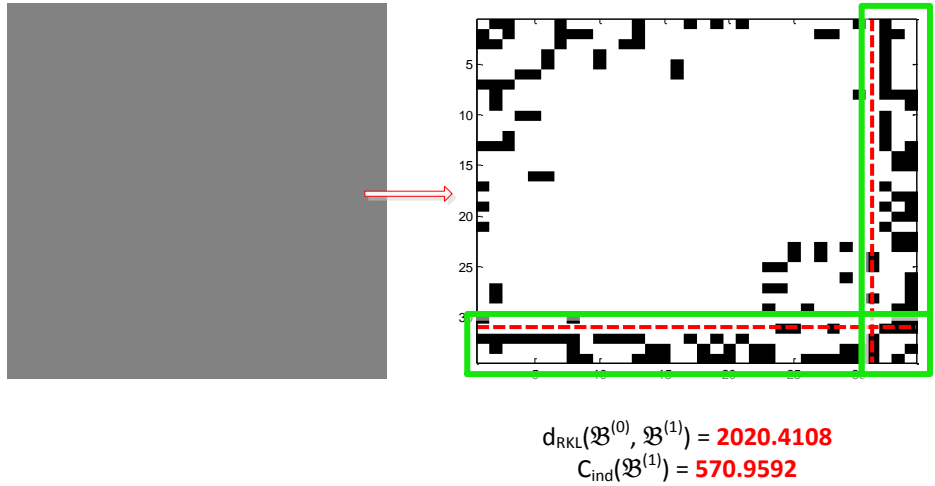

$d_{RKL}(\mathcal{B}^{(0)}, \mathcal{B}^{(1)})$ = **2020.4108**
$C_{ind}(\mathcal{B}^{(1)})$ = **570.9592**

(B) β = 5

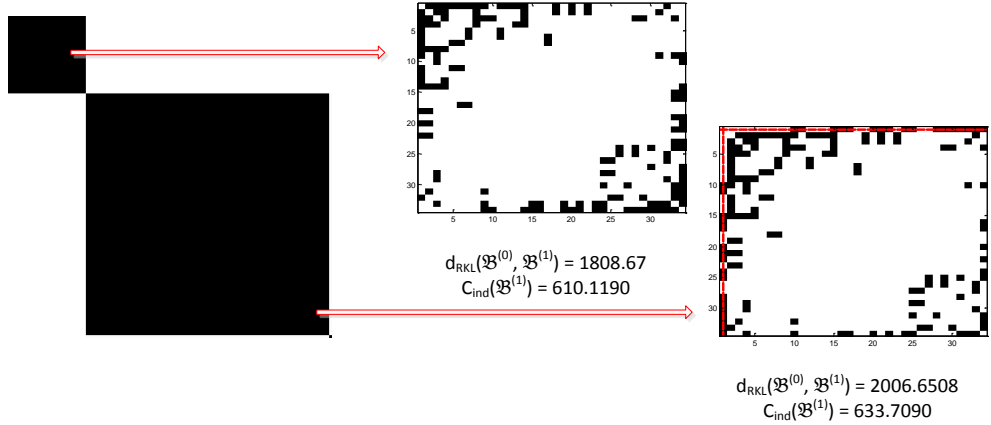

$d_{RKL}(\mathcal{B}^{(0)}, \mathcal{B}^{(1)})$ = 1808.67
$C_{ind}(\mathcal{B}^{(1)})$ = 610.1190

$d_{RKL}(\mathcal{B}^{(0)}, \mathcal{B}^{(1)})$ = 2006.6508
$C_{ind}(\mathcal{B}^{(1)})$ = 633.7090

(C) β = 10

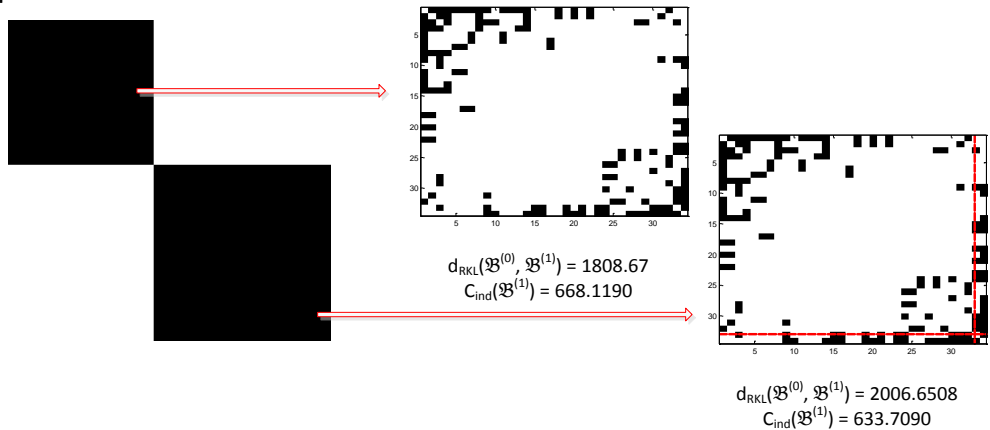

$d_{RKL}(\mathcal{B}^{(0)}, \mathcal{B}^{(1)})$ = 1808.67
$C_{ind}(\mathcal{B}^{(1)})$ = 668.1190

$d_{RKL}(\mathcal{B}^{(0)}, \mathcal{B}^{(1)})$ = 2006.6508
$C_{ind}(\mathcal{B}^{(1)})$ = 633.7090

**Figure 4-3: Alternative Blockmodels for *Zachary's karate club* dataset.**

Alternative blockmodels for different values of β parameter after running our second approach 100 times. An *iVAT* plot helps assess the cluster tendency. $d_{RKL}(\mathcal{B}^{(0)}, \mathcal{B}^{(1)})$ corresponds to the *Edge Reconstruction Kullback-Leibler Distance* between the given blockmodel $\mathcal{B}^{(0)}$ and the one discovered $\mathcal{B}^{(1)}$ for the corresponding β parameter. $C_{ind}(\mathcal{B}^{(1)})$ corresponds to the *Individual Snapshot Encoding* of $\mathcal{B}^{(1)}$. Both measurements are shown next to each alternative blockmodel. **(A) β = 1**. iVAT plot shows consistency across 100 runs. Just one alternative blockmodel can be discovered whose $d_{RKL}(\mathcal{B}^{(0)}, \mathcal{B}^{(1)})$ is the highest and $C_{ind}(\mathcal{B}^{(1)})$ is the lowest. **(B) β = 5**. iVAT plot shows that while β increases, a second blockmodel tends to occur more frequently. **(C) β = 10**. The alternative blockmodels are very similar to the ones with β = 5. It seems the cluster tendency will not change.

In Figure 4-3, *iVAT* helps assess the cluster tendency within the 100 runs for each β value. The *Edge Reconstruction Kullback Leibler* distance $d_{RKL}$ (See section 3.5) between the given blockmodel $\mathcal{B}^{(0)}$ and the one discovered for the corresponding $\beta$ parameter, that is $\mathcal{B}^{(1)}$, is shown next to each alternative blockmodel solution. Besides to the dissimilarity measurement, the quality measurement *Individual Encoding Cost* is also presented. For $\beta = 1$ just one alternative blockmodel is discovered. It happens that this is the chosen alternative blockmodel since it has the highest $d_{RKL}$ value and the lowest $C_{ind}$.

Now, let's find out if this alternative blockmodel representation makes sense. First of all, **the chosen representation resembles a core-periphery network structure** where the nodes of one position –core- are linked with many nodes of the same and/or other positions. Figure 4-4 shows the network division corresponding to the chosen alternative blockmodel. Core nodes are depicted with red colour whereas periphery with blue.

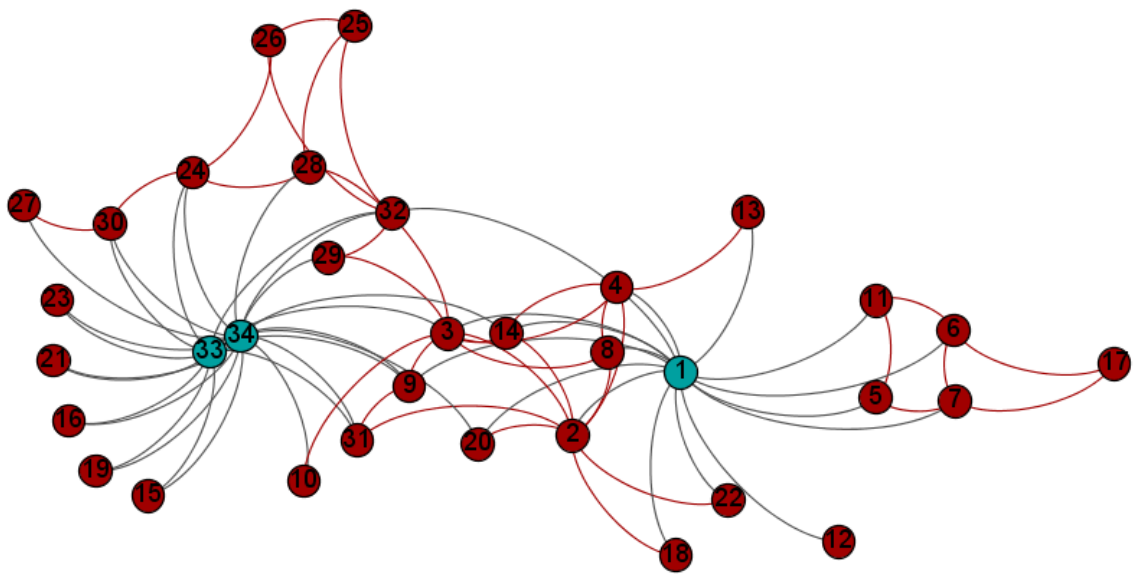

**Figure 4-4: Chosen alternative network division for *Zachary's karate club*.**

Blue nodes correspond to the *core* position whereas red nodes to the periphery.

Figure 4-5 shows how our second approach was able to find an alternative blockmodel solution which corresponds to a different network structure.

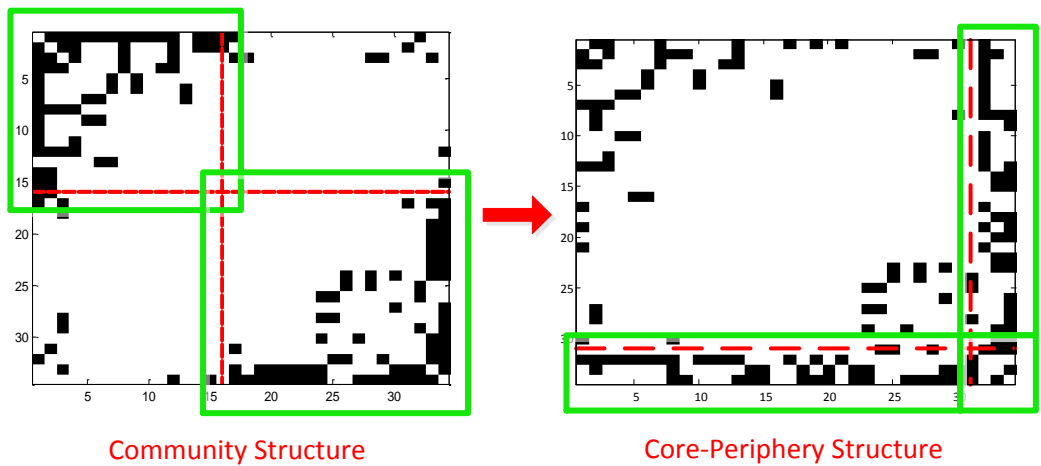

Community Structure          Core-Periphery Structure

$$M^{(0)} = \begin{pmatrix} 0.4643 & 2.1947e-24 \\ 2.1931e-24 & 0.4508 \end{pmatrix} \quad\quad M^{(1)} = \begin{pmatrix} 7.4898e-86 & 0.6521 \\ 0.6521 & 1.2794e-51 \end{pmatrix}$$

**Figure 4-5: Core-periphery alternative blockmodel solution for *Zachary's karate club*.**

The image matrix $M^{(1)}$ of the core-periphery structure is totally different from $M^{(0)}$ since the algorithm maximised the dissimilarity between image matrices (See section 3.4.)

We can even regard the alternative blockmodel shown at the right part in iVAT and blockmodels for different $\beta$ parameter values. Note the much higher value for $\beta$ with respect to the value needed when the discovering process was in the other way (Compare with Figure 4-3.)

Figure 4-5 as the initial blockmodel $\mathfrak{B}^{(0)}$ and try to discover the original blockmodel presented in Figure 4-2 as its alternative solution.

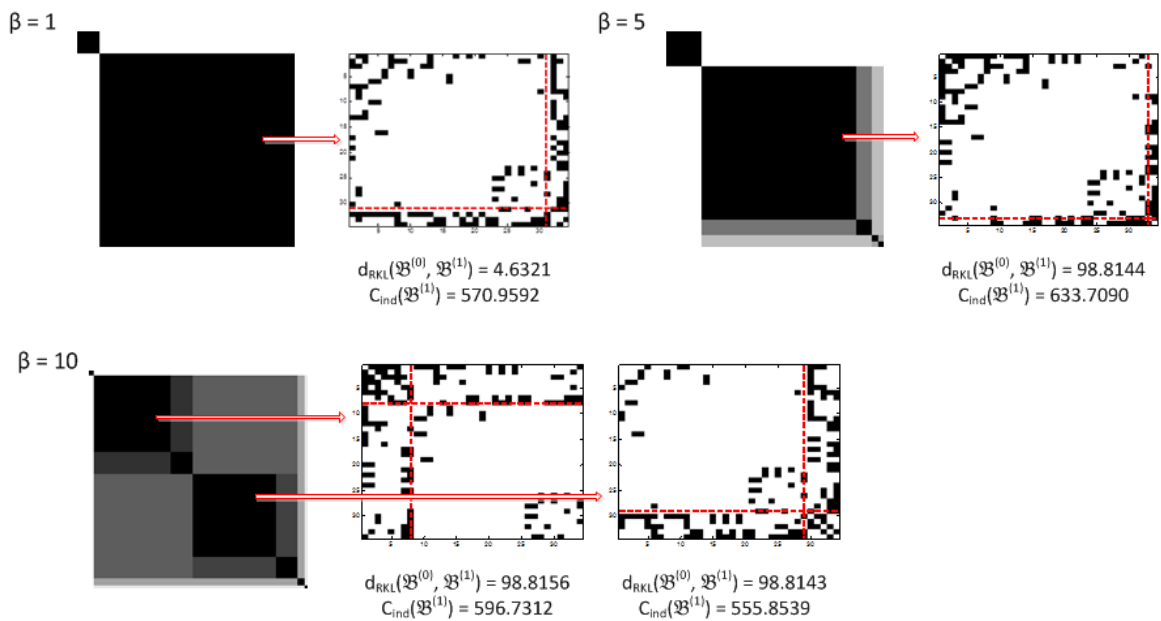

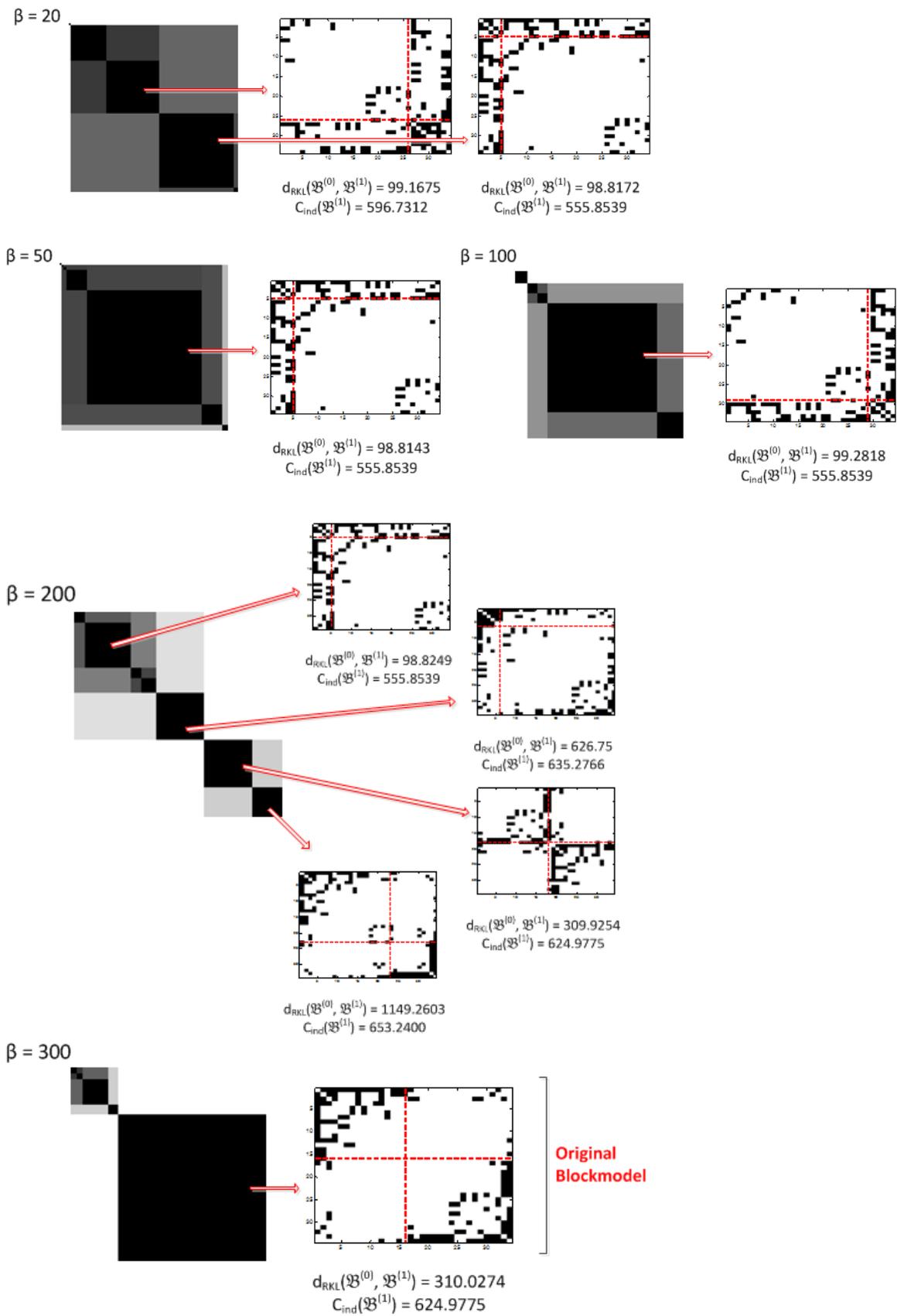

**Figure 4-6: Original *karate club's* blockmodel discovered as alternative.**

Figure 4-6 shows the process of discovering an alternative blockmodel when the given blockmodel $\mathfrak{B}^{(0)}$ is the alternative solution found previously. It is interesting to see the value $\beta$ parameter took to discover the original blockmodel. This $\beta$ value is much higher than the $\beta$ value needed in the other way. One possible explanation is that our approach actually tends to find a solution which is different to the actual division of this particular dataset. Then, the dissimilarity parameter does not need to be big ($\beta = 1$), whereas if the approach is forced to find a very different solution to the one it is used to, it demands much more effort ($\beta = 300$).

On the other hand, the behaviour of the dissimilarity measure $d_{RKL}$ confirms the effect of the trade-off parameter on the whole objective function. The bigger the trade-off parameter is set –which means more effect of the dissimilarity term-, the bigger the value of the measurement is. However, there is a kind of plateau that spans from small to big values of $\beta$ where $d_{RKL}$ is roughly equal to 100. It can be thought that the objective function reached a local optimum from which it could jump off somehow when the trade-off parameter increased.

Now, let's have a look at another real dataset which refers to *Books on American politics*. The nodes within this dataset can be clustered according to their political alignment, i.e. liberal, conservative or centrist. In fact, in the original dataset every book came classified as one of those tendencies. Figure 4-7 shows the actual division accordingly.

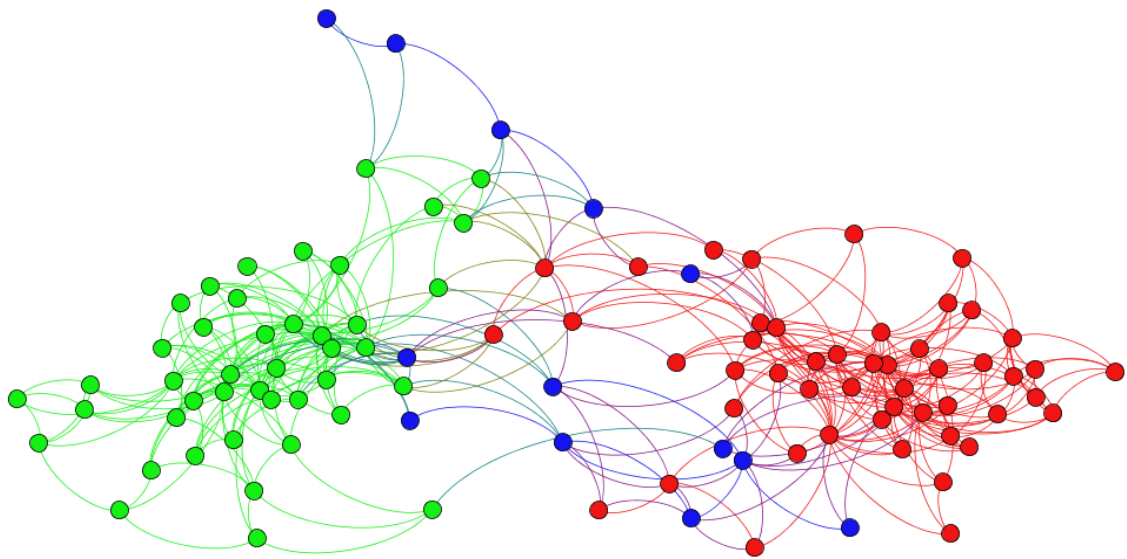

**Figure 4-7: Actual division of books on American politics dataset.**

In this figure, liberal books are depicted with green, centrist with blue and conservative with red. *Any resemblance to reality is pure coincidental*.

Therefore, the initial blockmodel $\mathfrak{B}^{(0)}$ with its corresponding image matrix $M^{(0)}$ are shown in Figure 4-8.

**(A)**

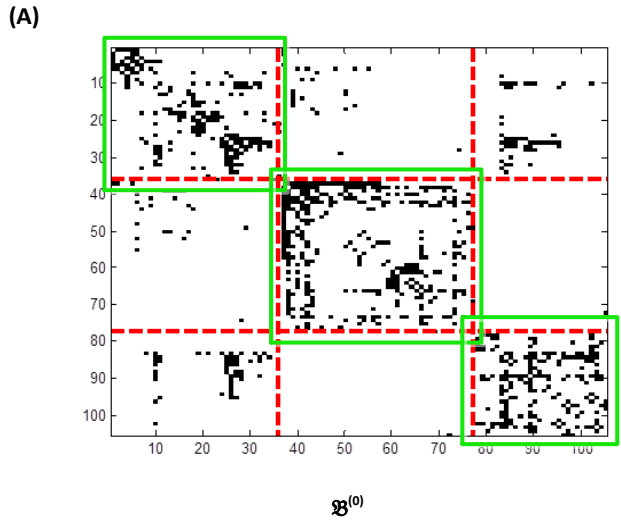

$\mathcal{B}^{(0)}$

**(B)**

$$M^{(0)} = \begin{pmatrix} 0.2759 & 3.21e-63 & 2.20e-27 \\ 3.59e-63 & 0.3581 & 0 \\ 1.80e-28 & 0 & 0.4514 \end{pmatrix}$$

**Figure 4-8: Blockmodel representation of the actual division of books on American politics.**

(A) A blockmodel representation of the community structure within the dataset.

(B) Image matrix $M^{(0)}$ used as input in our second approach.

Our second approach is run over the image matrix $M^{(0)}$ of Figure 4-8 (B). Figure 4-9 shows alternative blockmodels for different values of $\beta$ parameter. In this figure, we have dropped *iVAT* plots due to space reasons. However, in some cases, we have chosen more than one alternative solution for an individual parameter value since their occurrence within the 100 runs is similar –this would have been more obvious by observing *iVAT plots*.

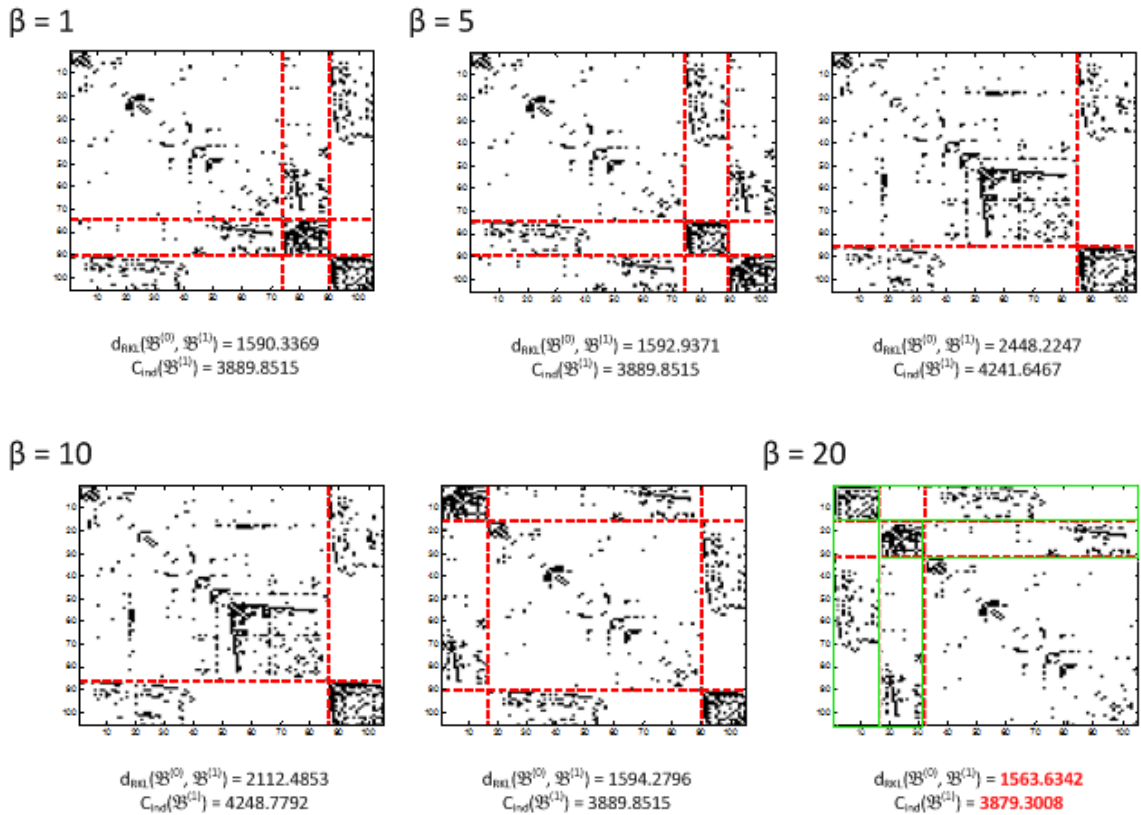

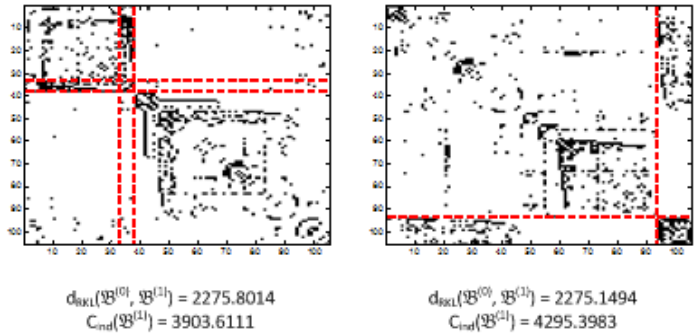

Figure 4-9: Alternative blockmodels for *Books on American politics* dataset.

*iVAT* plots have been dropped, yet, in some cases, more than one alternative blockmodel is shown for each $\beta$ parameter. At the bottom of each blockmodel, $d_{RKL}$ and $C_{ind}$ are also presented. The chosen blockmodel is the one resulting when $\beta = 20$.

From the alternative blockmodels in Figure 4-9, we have chosen the one with better quality ($\beta = 20$) which shows an interesting core-periphery structure of two tiers (See Figure 4-10.)

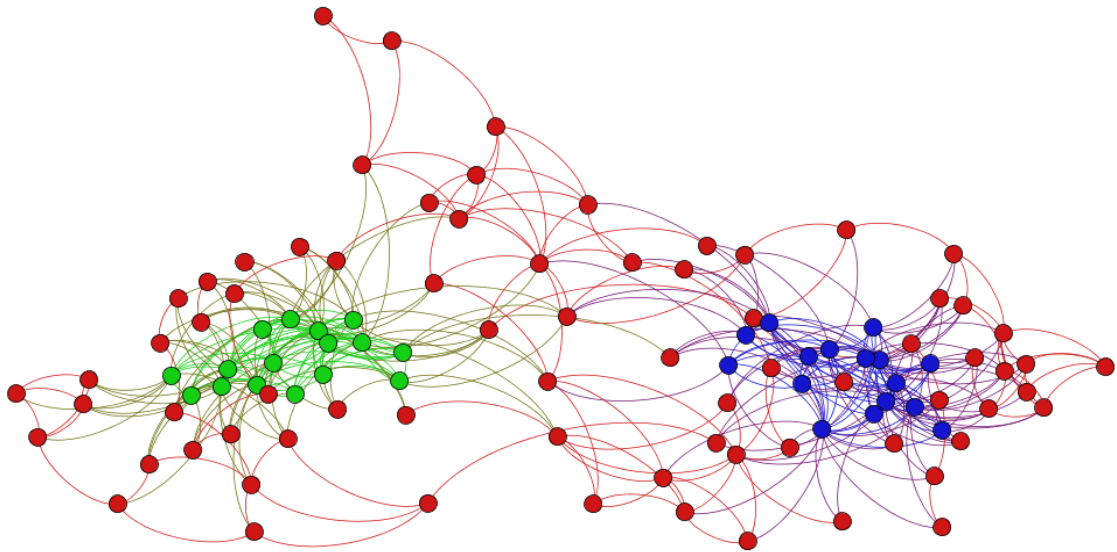

Figure 4-10: Chosen alternative network division for *Books on American politics.*

In this figure, the cores of the liberal and conservative ideologies are depicted with green and blue respectively, whereas the periphery nodes are shown in red.

Again, a core-periphery structure is discovered given a community configuration as input. In both datasets, the algorithm is prone to discover dense positions with a small number of nodes. In other words, these positions contain the nodes with the highest degree within the network which leads to the core-periphery network structures.

Furthermore, it is also interesting to see that some alternative blockmodel solutions present less number of positions with respect to the reference blockmodel. In Figure 4-9, $\beta = 5, \beta = 10$ and $\beta = 50$ show blockmodels with two positions. In these cases, the image matrix keeps its original dimensions, i.e. 3 x 3, yet the position membership matrix presents a column vector which has the smallest values among the columns for every row.

## 4.2. Approaches to solve this problem

In the previous section, in order to present the problem, we have showed experiments by using exclusively our second approach. Therefore, in this section, we are going to focus on making comparisons between both approaches by applying them on a synthetic dataset.

**This synthetic dataset was constructed by mixing two pre-defined clusterings.** Each clustering has three positions, yet their network structure is different. The first clustering has a well-defined community structure, whereas the second is core-periphery.

Figure 4-11 shows at the left the original weighted adjacency matrix. It has 50 nodes and the edge occurrence is sampled from a binomial distribution and the weights were from a lognormal distribution. At the right side, the figure shows the constructed clusterings which are sought through the application of our both approaches. Nevertheless, it does not mean that the optimisation process will not be able to find other alternatives. In fact, **other plausible solutions were discovered by our approaches** as it will be shown below.

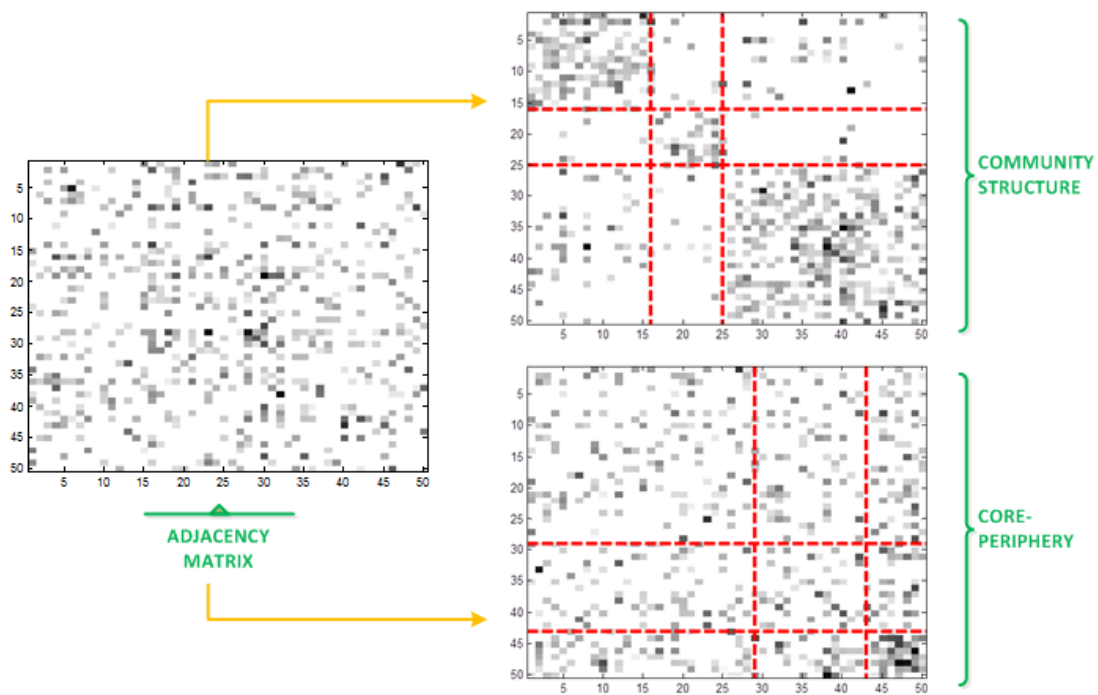

Figure 4-11: Original Adjacency Matrix which was created from mixing two blockmodel structures.

The adjacency matrix has 50 nodes and the weights of the edges were sampled from a lognormal distribution with mean = 5 and standard deviation = 0.5. An edge between two nodes corresponds to a successful occurrence within a binomial distribution of $|C_r| * |C_c|$ elements and probability equal to a pre-defined density for the block (block resulting from the intersection of the positions $C_r$ and $C_c$).

Our first set of experiments within this section regards the **community structure blockmodel as the given information**. Thus, our goal is to eventually find the core-periphery structure as an alternative blockmodel.

Both approaches depend on a user parameter $\beta$ which controls either the effect of the *cannot-link* constraints matrix or the effect of the dissimilarity between image matrices on the whole objective function. This effect is illustrated from four different perspectives.

**It is important to mention that the magnitude of the $\beta$ parameter is different for each approach. Hence, it is not correct to compare the approaches based on it. However, we have experimentally defined the ranges of this parameter for each approach so that plausible alternative blockmodels were found within them.**

For each $\beta$ value we have run each approach 100 times. **The most occurring alternative blockmodel is chosen as representative for that $\beta$ value.** The four perspectives from which this chosen alternative blockmodel is evaluated are:

a) *Quality: Individual Snapshot Encoding Cost $C_{ind}$* is calculated for the chosen alternative blockmodel.
b) *Dissimilarity: Kullback-Leibler Reconstruction Measure $d_{RKL}$* with respect to the given blockmodel.
c) *Similarity to a target:* Recall that our goal is to eventually find the core-periphery or community structure as an alternative blockmodel with respect to a given community or core-periphery structure, respectively. Although we have discussed about the inaccuracy of measures like *NMI* when evaluating similarity between blockmodels, we use it since the familiarity of researchers with this metric.
d) *Occurrence:* The number of times out of 100 runs the chosen alternative blockmodel appeared.

Figure 4-12 shows that our second approach has alternative blockmodels with encoding costs less than 1700, whereas the minimum cost in our first approach is not less than 1850 as the first measure cannot be taken into account since it was with $\beta = 0$. Furthermore, the first approach has measures close to 1950 which does not happen with approach 2. It is also interesting to see that "best" alternative solution for the second approach which, despite of keeping a community structure, corresponds to a plausible solution as one of the communities in the alternative blockmodel brings together two communities of the given one. In other words, **although we could not find the target alternative blockmodel –core-periphery-, we were able to find a good quality alternative blockmodel.**

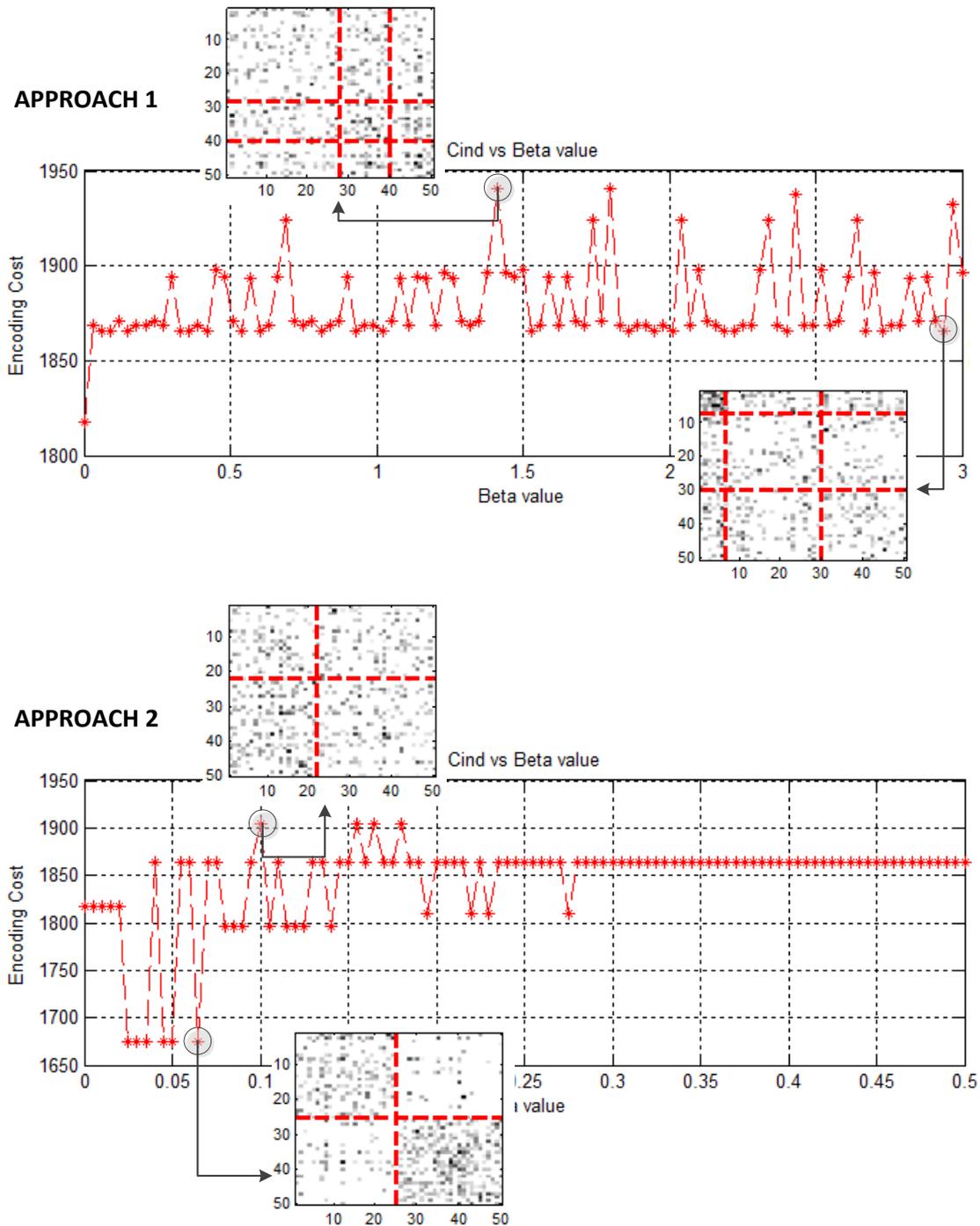

Figure 4-12: Quality measurements with different $\beta$ values for both approaches.

The scenario is different for dissimilarity. Now, approach 1 shows a sort of plateau of 0.65, whereas for approach 2 there are even measures of dissimilarity equal to 0. Our second approach also shows several measures less than 0.4. Thus, lack of better quality alternative blockmodels in approach 1 is compensated by more dissimilar solutions in this particular situation (See Figure 4-13.)

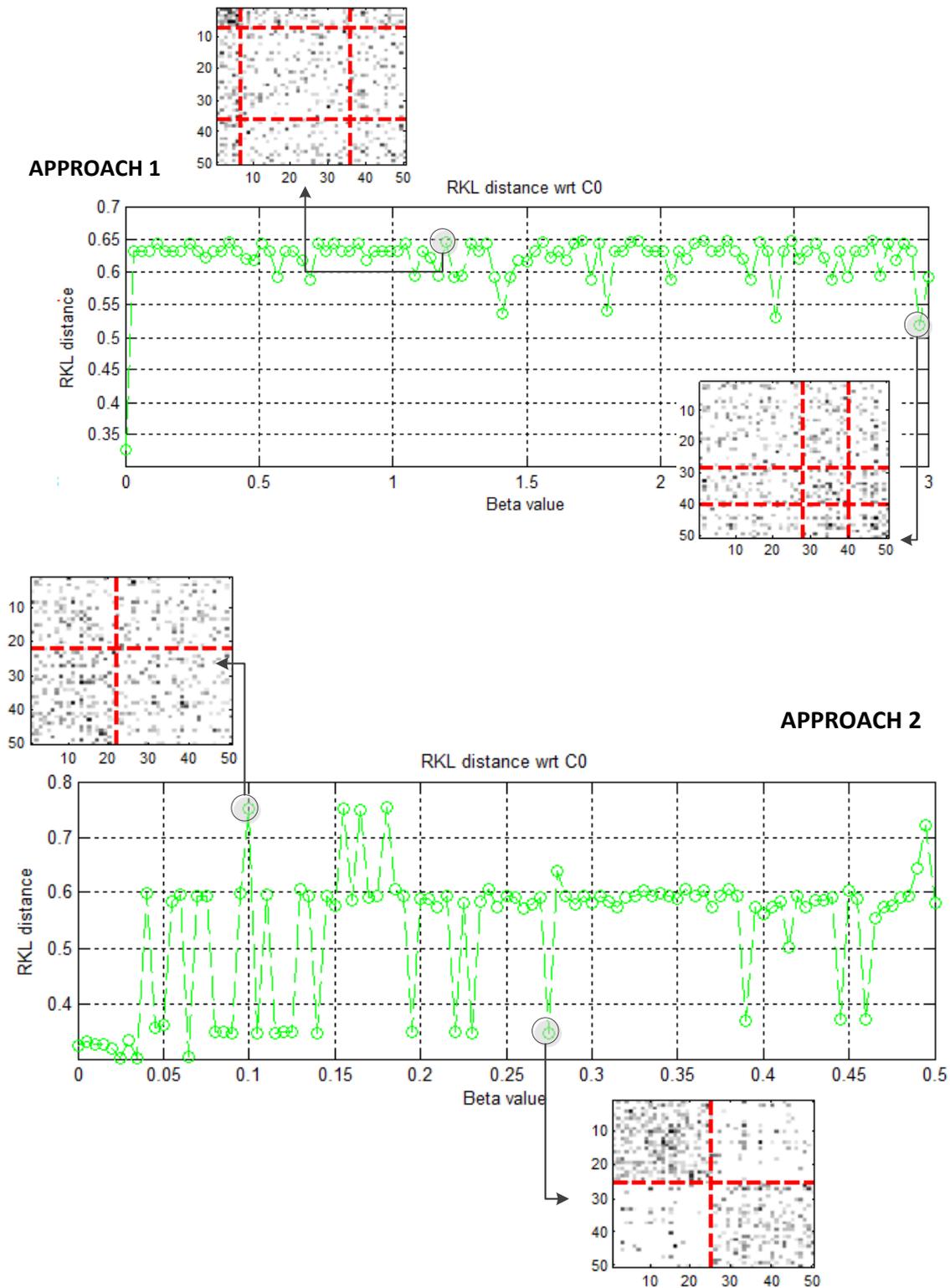

**Figure 4-13: Dissimilarity measurements with different $\beta$ values for both approaches.**

The third perspective is the similarity with respect to the target blockmodel. Since we ran the approaches regarding that the given clustering is the one with the community structure, our target becomes the core-periphery. We show how close/far we are from this goal by *NMI*.

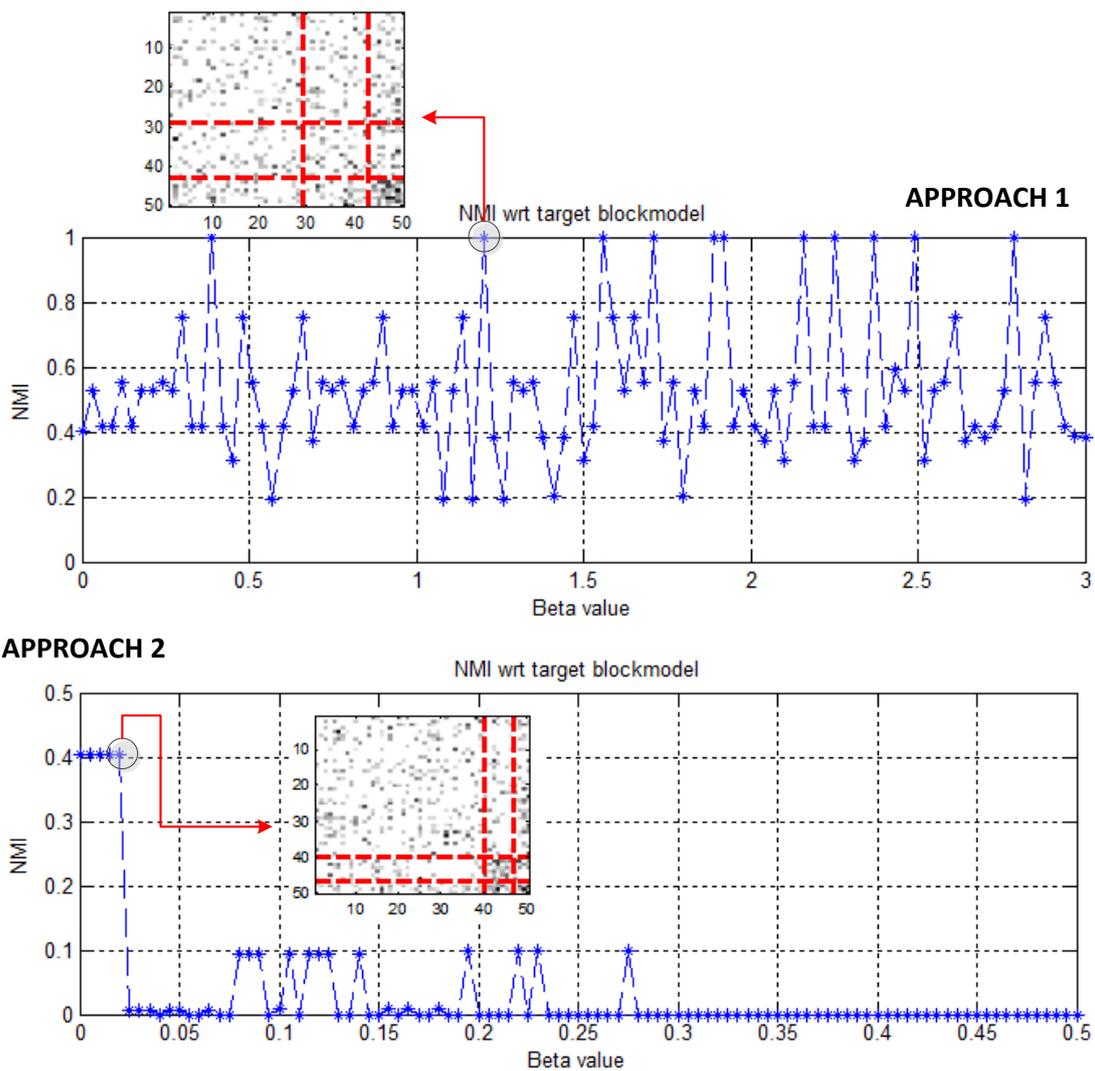

Figure 4-14: Similarity measures with respect to the core-periphery blockmodel structure with different $\beta$ values for both approaches.

From this perspective, our first approach shows a much better performance. Approach 2 roughly reaches 0.4 of similarity with respect to the core-periphery structure while approach 1 could discover it several times. **Nonetheless, the alternative blockmodel found by our second approach ($NMI \sim 0.4$,) perhaps presents a better core-periphery structure than the target** (See Figure 4-14.)

The fourth perspective tries to show how likely an alternative solution could be for each approach with different values of $\beta$. There are two noticeable facts shown in Figure 4-15: *(a)* the average number of occurrences in our first approach is much lower than in our second approach; and *(b)* there is a positive linear association between this number and $\beta$ in approach 2 which has not been clearly observed in the other perspectives. Then, **it is necessary to have enough evidence to reject the null hypothesis of no effect of this parameter within both approaches**.

**APPROACH 1**

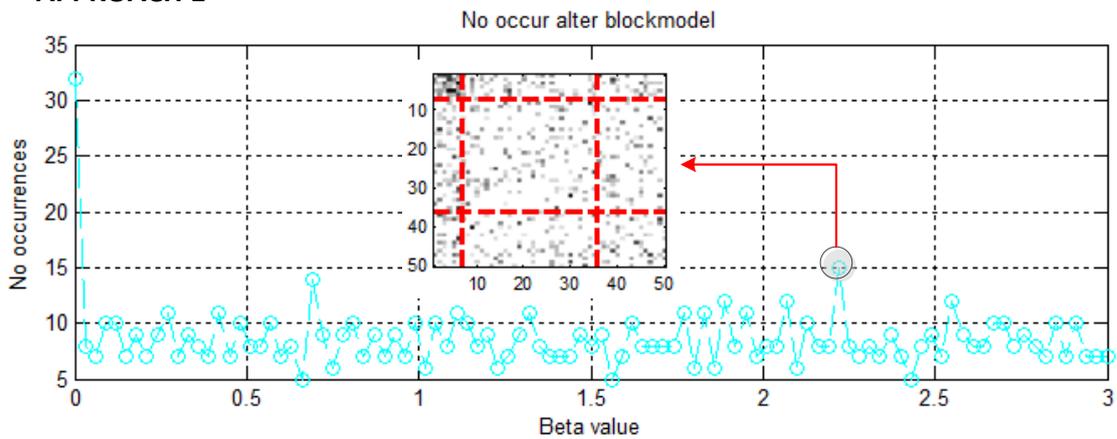

**APPROACH 2**

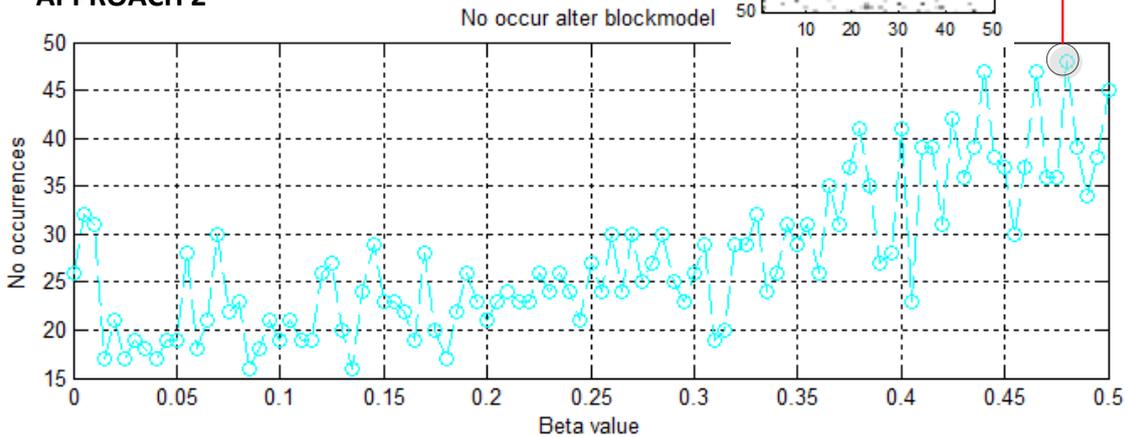

Figure 4-15: Number of occurrences of the most common alternative blockmodel with different $\beta$ values for both approaches.

### 4.2.1. Effect of $\beta$ parameter on both approaches

In order to conclude with respect to the influence that $\beta$ has on both approaches, we performed analysis of variance –*ANOVA*- of the linear model of the variables. The metrics for the mentioned perspectives –quality, dissimilarity, similarity w.r.t. the target and number of occurrences- are regarded as dependent variables, whereas $\beta$ corresponds to the explanatory variable. In this study we have taken the 100 observations already made in the previous section for each approach. We have also adopted the widely used threshold for *P-value* < 0.05 to consider a statistic as significant –statistically significant[21].

| Approach | P-value | | | |
|---|---|---|---|---|
| | $C_{ind}$ w.r.t. given blockmodel | $d_{RKL}$ w.r.t. given blockmodel | $NMI$ w.r.t. target blockmodel | No. occurrences most common alternative blockmodel |
| Instance-level constraints | 0.5333 | 0.3832 | 0.7216 | 0.05988 . |
| Dissimilarity between image matrices | 3.64e-05 *** | 0.0008532 *** | 0.0003873 *** | 2.2e-16 *** |
| | Signif. codes:  0 '***' 0.001 '**' 0.01 '*' 0.05 '.' 0.1 ' ' 1 | | | |

Table 3: *P-value* from *ANOVA* of the linear models of the variables $\beta$ (explanatory) vs. metrics (response)

From Table 3, we can conclude that **there is strong evidence to reject the null hypothesis of no effect of the trade-off parameter only for our second approach. Our first approach is not sensitive to $\beta$. This conclusion implies that our metrics, in the first approach, are correlated with a confounding variable which is the number of constraints.** Therefore, the next section explains the level of influence of this variable on the performance of approach 1.

### 4.3. Considerations about instance-level constraints

The experiments carried out in the previous section for approach 1 used a "special" *cannot-link* constraints matrix. What we mean with special is that this matrix contains the **perfect** set of pairs of constraints. In this set, every constraint was constructed taking into account not only the given blockmodel but also the target. For example, if nodes *i* and *j* belong to the same position in the given blockmodel but not in the target, a constraint is constructed; otherwise, not. This procedure has three **properties**:

a) *Every constraint is correct*. It tries to break a link between two nodes in the given blockmodel if and only if there is not a link between them in the target. Sadly, the likelihood of constructing an incorrect constraint is very high since we do not know the target beforehand.

b) *The clustering process is feasible.* Although the feasibility problem for *cannot-link* constraints is **NP**-complete [40], knowing the target allows constructing a set of constraints which makes the clustering process feasible. In other words, there exists a partition of the network which satisfies the whole set of *cannot-link* constraints which is evidently the target.

---

[21] *P-value* is the probability of observing a value of a statistic –in this case: *F* statistic- as or more extreme than the one actually observed, assuming that the null hypothesis is true. Recall that the null hypothesis corresponds to the case of no effect of the trade-off parameter over the metrics.

c) *The number of constraints is appropriate.* We show below how important the role of the number of constraints is in guiding the clustering process.

However, the target blockmodel is unknown beforehand, or at least we cannot count on knowing the whole structure of a target solution. Nonetheless, some information besides to the given clustering might be unveiled. Therefore, instead of using the whole perfect set of constraints, we experiment with a gradually changed percentage of them. In this way we are keeping properties *(a)* and *(b)* which are quite important as we will show below.

Now, let's see what the behaviour of approach 1 is when only a portion of correct *cannot-link* constraints is provided. The following experiments also include the trade-off $\beta$ parameter in order to see if both variables **interact** with each other. Thus, the experiments consist of the observations of the four metrics for each of the 180 combinations of the percentage of correct constraints and $\beta$ parameter –10 values of the percentage and 18 values of $\beta$.

Multiple-regression models were fit to these observations and *ANOVA* was performed on them to discover whether the interaction of the factors is statistically significant.

|  | $C_{ind}$ w.r.t. given blockmodel | $d_{RKL}$ w.r.t. given blockmodel | NMI w.r.t. target blockmodel | No. occurrences most common alternative blockmodel |
|---|---|---|---|---|
| % correct constraints | **8.569e-08 \*\*\*** | **<2e-16 \*\*\*** | **< 2e-16 \*\*\*** | **<2e-16 \*\*\*** |
| $\beta$ parameter | 0.2469 | 0.2675 | 0.34124 | 0.2178 |
| Interaction of both factors | 0.5282 | 0.7416 | **0.03861 \*** | 0.3916 |

```
Signif. codes:  0 '***' 0.001 '**' 0.01 '*' 0.05 '.' 0.1 ' ' 1
```

Table 4: *P-value* from *ANOVA* of the multiple-regression models of the variables $\beta$ and percentage of correct constraints (explanatory) vs. metrics (response)

From Table 4, **we can conclude that we do not have enough evidence to reject the null hypothesis of no interaction between the factor variables to explain the variation of the quality, dissimilarity and number of occurrences of the most common blockmodel. As concluded in the previous section, the main effect of $\beta$ parameter over the variation of the metrics is not statistically significant. Finally, the percentage of correct constraints do influence on the dependent variables. The greater the number of correct constraints, the greater the possibility of finding a good alternative blockmodel.**

Therefore, **the existence of correct side information about any alternative blockmodel would improve the performance of our first approach dramatically. It is obvious we cannot count on the whole target, yet any information would be quite useful**.

# Chapter 5

# Discussion and Conclusions

This chapter summarises the most important findings in this work and also analyses the usage of instance-level constraints with negative connotation.

From Chapter 2 we can conclude that the relevant work within the field of alternative clustering only considers non-relational data. When the dataset represents a network, only community structures have been sought. There are not attempts to summarise the structure of a network through alternative blockmodels based on given information, that is, initial blockmodels. Therefore, our two approaches are a novel contribution within both graph clustering and blockmodelling fields.

Our approaches presented in Chapter 3 are based on two important techniques: *(a)* SNMTF, and *(b) Iterative Lagrangian Solution* [12]. *SNMTF* has been proved to be an effective clustering tool within undirected graphs and it also fits nicely with the blockmodelling problem. In turn, the *Iterative Lagrangian Solution* is a convenient normalisation technique which follows the stochastic nature of the position membership matrix and avoids the uncertainty caused when the factor matrices –position membership and image matrices- have different magnitudes.

The inclusion of *cannot-link* constraints to guide the clustering process has been shown to be a powerful tool but very sensitive to the number and quality of constraints (See Chapter 4.) We have enough evidence to conclude that the number of constraints has a statistically significant effect on the values of the evaluation metrics. Furthermore, when experiments regarding "incorrect" constraints were carried out, the evaluation metrics were not satisfactory[22].

Incorrect constraints were constructed taking into account no target information, that is, they only considered the given blockmodel. The quality of the constraints is dubious because there is no guarantee that at least one feasible division of the network exists that holds all the constraints. For instance, if the *cannot-link* constraints $c_1(a,b)$, $c_2(b,c)$ and $c_3(a,c)$ were created as $a, b\ and\ c$ belong to the same position in the given blockmodel, there is no feasible alternative blockmodel with three positions that holds these constraints. What would it have happened if we would know that $a\ and\ c$ belong to the same position in the target alternative blockmodel beforehand? The answer is that $c_3$ would not have been created and the clustering process would be feasible. Therefore, side information of the target alternative blockmodel is quite valuable for our first approach.

Experiments in Chapter 4 also showed that only our second approach is sensitive to the trade-off parameter $\beta$. Moreover, there is no interaction between factors: trade-off parameter and number of constraints.

---

[22] These experiments are not included in Chapter 4 because their results did not show any pattern.

On the other hand, our second approach does not have to deal with the complexity of construction and identification of correct constraints. It is a single-parameter approach which is able to find a good quality alternative blockmodel. However, it does not seem to be as powerful as our first approach as it was shown in Chapter 4.

Both approaches were able to find dissimilar and good quality alternative blockmodels with a common characteristic: they are prone to find "small" positions. In other words, they tend to discover positions with nodes that have lots of connections, i.e. central nodes. It means that core-periphery structures are very likely to be found. This trait is inherited from the *Iterative Lagrangian* technique as it was discovered while we were in the experimental stage.

We also found empirically that the inclusion of *must-link* constraints made the clustering process harder. Identification of "correct" *must-link* constraints and their number had the same difficulties as with *cannot-link* constraints. Moreover, when they were included, randomness of the alternative solutions increased.

Future work is needed for identification and construction of instance-level constraints as they are an effective tool in guiding the clustering process. Unfeasibility due to over-constraining and complexity in the construction are the main disadvantages when dealing with cannot-link constraints.